\documentclass[twocolumn]{aastex631}

\usepackage{CJK}
\usepackage{tabularx}

\shorttitle{LMC X-4}
\shortauthors{Chou}

\begin{document}
\begin{CJK*}{Bg5}{bsmi}
\title{Superorbital Phase Evolution and a Soft-Hard X-ray Phase Shift in LMC X-4 }

\correspondingauthor{Yi Chou}
\email{yichou@astro.ncu.edu.tw}

\author[0000-0002-8584-2092]{Yi Chou (©PÖö)}
\affiliation{Graduate Institute of Astronomy, National Central University \\
  300 Jhongda Rd. Jhongli Dist. Taoyuan, 320317, Taiwan}



\begin{abstract}

The superorbital period of LMC X-4 is among the most stable known in Roche-lobe overflow, high-mass X-ray binaries. We analyzed 33 years of monitoring data from the Compton Gamma Ray Observatory Burst and Transient Source Experiment (CGRO BATSE), the Rossi X-ray Timing Explorer All-Sky Monitor (RXTE ASM), the Neil Gehrels Swift Burst Alert Telescope (Swift BAT), the Monitor of All-sky X-ray Image Gas Slit Camera (MAXI GSC), and the Fermi Gamma-ray Burst Monitor (Fermi GBM). The measured phases show a smooth long-term trend with superposed systematic fluctuations. Fits with cubic, quartic, and sinusoidal models indicate that the quartic and sinusoidal forms provide significantly better descriptions, with the sinusoidal model yielding an $8900^{+210}_{-230}$-day modulation. Such a long timescale is unlikely to arise from orbital motion around a tertiary companion. The fluctuations resemble stochastic, glitch-like events on several-hundred-day timescales. Their rms period variation exceeds that of the smooth trend, yet the total rms period variation over 33 years remains only 0.55\%, demonstrating the exceptional stability of the superorbital period. During MJD 57000-60461, we detect a phase offset of 0.044$\pm$0.010 cycles between the soft and hard X-ray bands. This offset can be reproduced by including a higher-harmonic term in the azimuthal disk model, allowing a transition from antisymmetric to asymmetric structure. A contemporaneous decline in the hard X-ray flux suggests a partial obscuration of the emission region, similar to the anomalous low state in Her X-1. This evolving-disk scenario may also explain the superorbital phase shift previously reported in Her X-1.

\end{abstract}


\section{Introduction} \label{intro}

LMC X-4, first identified by~\citet{gia72}, is a high-mass X-ray binary (HMXB) in the Large Magellanic Cloud (LMC). The system consists of a 1.57 $M_{\sun}$ neutron star and a mass-losing O8\, {\footnotesize III} companion of 18 $M_{\sun}$~\citep{fal15}. LMC X-4 is an eclipsing system with an orbital period of 1.4 days~\citep{li78} and exhibits pulsations with a spin period of 13.5 s~\citep{kel83a}. Initial evidence for an orbital-period derivative of ${\dot P}{\mathrm{orb}} / P{\mathrm{orb}} = (1.1 \pm 0.8) \times 10^{-6}$ $\mathrm{yr}^{-1}$ was reported by~\citet{lev91}, but subsequent observations yielded negative values~\citep{saf96,woo96,lev00,pau02,mol15,fal15}. The most recent measurement by~\citet{jai24} reports an orbital period derivative of ${\dot P}{\mathrm{orb}} / P{\mathrm{orb}} = (-1.287 \pm 0.010) \times 10^{-6}$ $\mathrm{yr}^{-1}$, accompanied by a significant second derivative of ${\ddot P}_{\mathrm{orb}} = (-2.5 \pm 0.4) \times 10^{-12}$ $\mathrm{day}$ $\mathrm{day}^{-2}$. Such a large orbital period derivative is attributed to tidal torques arising from asynchronism between the orbital motion and the companion's rotation~\citep{kel83b,lev00}. Conversely, the 13.5 s pulsation, originally discovered by~\citet{kel83a}, has been confirmed by various X-ray observatories~\citep{nar85,pie85,lev91,woo96,pau02,nai04,hun10,mol17,bru20}. By combining historical records with spin period measurements from RXTE, XMM-Newton, and the X-ray telescope (XRT) on board the Neil Gehrels Swift Observatory (hereafter Swift),~\citet{mol17} found that the spin period shows a near-periodic variation with a timescale of 6.8 years and an amplitude of $7 \times 10^{-4}$ s. Such a periodic variation is more plausibly attributed to torque reversals than to the binary system orbiting a massive hierarchical third companion, because the measured radial velocities of LMC X-4 remain consistently around 300 km s$^{-1}$ across different epochs, with fluctuations far smaller than those expected from orbital Doppler shifts induced by a massive third companion~\citep{mol17}.

In addition to pulsations and eclipses, LMC X-4 demonstrates a superorbital modulation with a period of 30.5 days, substantially exceeding its 1.4-day orbital period. This long-term variability was initially identified by~\citet{lan81} in the 13-80 keV X-ray band using HEAO-1 observations spanning $\sim$500 days, during which the X-ray flux was observed to fluctuate by up to a factor of 60. Subsequent power spectral analysis of the optical counterpart by ~\citet{ilo84} further corroborated the presence of this superorbital signal. Later monitoring using several X-ray monitoring/scanning instruments, including CGRO BATSE~\citep{mol15}, Ginga ASM~\citep{pau02}, RXTE ASM~\citep{wen06}, Swift BAT~\citep{mol15}, and MAXI~\citep{hu15,mol15} validated the persistence of the modulation. The persistence of this 30.5-day modulation is generally attributed to a warped, precessing accretion disk whose precession is maintained by torques acting on the disk.

\citet{lan81} and \citet{ilo84} posited that the superorbital variability originated from the periodic occultation of the central X-ray source and the inner disk by a precessing tilted accretion disk. ~\citet{hee89} refined this interpretation by demonstrating that the $\sim$30-day modulation of the optical orbital light curve can be reasonably explained by a geometric model that incorporates ellipsoidal variations, X-ray heating  of the companion, and the effects of a precessing tilted disk.  Alternatively, the radiation-driven warping model developed by~\citet{pet77,pri96,wij99} attributes the modulation to a stable warp in the accretion disk, driven by radiation pressure and precessing under radiation torques. This model successfully accounted for the superorbital modulations in SMC X-1, Her X-1, and LMC X-4, although~\citet{tro07} showed that SMC X-1 does not maintain a single stable precession cycle. Its superorbital period wanders between $\sim$40 and 60 days, with substantial cycle-to-cycle variations indicative of an evolving or unstable disk warp. More recent studies further report that SMC X-1 exhibits erratic behavior with its superorbital period ranging from  39 to 72 days depending on whether the source enters an excursion state~\citep{bru23,dag22}.

\citet{ogi01} investigated the stability of precession of a radiation-driven warped accretion disk. Their results indicated that LMC X-4 resides near the boundary of the stable region for mode-0 warping (refer to Figure 7 in~\citet{ogi01} or Figure 1 in~\citet{kot12}) assuming plausible values for the disk viscosity ($\alpha = 0.3$) and accretion efficiency ($\eta = 0.1$). The stability of the LMC X-4 superorbital period was validated by~\citet{cla03} and~\citet{kot12} using a dynamic power spectral analysis of the RXTE ASM light curve. Although~\citet{cla03} identified a potential period jitter of $\pm \sim 0.05$ days in their power spectrum, this variation was negligible compared with the 3$\sigma$ uncertainty of $\pm 0.46$ days.

Phase evolution analysis offers a more sensitive diagnostic of period variability. The evolution of the superorbital phase of LMC X-4 was reported by~\citet{pau02, cla03, hu15, mol15}.~\citet{pau02} combined superorbital phases measured with the EXOSAT ME, GINGA ASM, and RXTE ASM  reporting a superorbital period derivative of $-2 \times 10^{-5}$ day day$^{-1}$.~\citet{cla03} constructed a dynamic folded light curve using the first seven years of RXTE ASM data to investigate superorbital phase variability, however they observed no significant phase shift, establishing an upper limit of $\sim$0.1 cycles.~\citet{hu15} analyzed light curves from RXTE ASM, Swift BAT, and MAXI spanning $\sim$18.3 years, revealing that the phase evolution could be described by a quadratic trend with a period derivative of $(2.08 \pm 0.12) \times 10^{-5}$ day day$^{-1}$, superimposed on systematic residual phase fluctuations. Conversely,~\citet{mol15}, analyzing Swift BAT data from February 2005 to May 2015 found no statistically significant period variation during that interval, although they observed residual phase fluctuations analogous to those reported by~\citet{hu15}. To accommodate  these fluctuations,~\citet{mol15} developed an empirical ephemeris to apply to data from CGRO BATSE, RXTE ASM, Swift BAT, and MAXI but the underlying physical mechanism responsible for these systematic fluctuations remains unknown.

For pulsars embedded in warped accretion disks, such as LMC X-4, a phase offset between soft and hard X-ray pulses is observed. The spectra of many X-ray binary pulsars can be well described by a power-law component from the neutron star plus a soft excess. ~\citet{hic04} proposed that this soft excess arises from the reprocessing of hard X-rays by the inner accretion disk in systems with luminosities $L_{X} \gtrsim 10^{38}$ erg s$^{-1}$, such as SMC X-1 ($L_{X} = 2.4 \times 10^{38}$ erg s$^{-1}$) and LMC X-4 ($L_{X} = 1.2 \times 10^{38}$ erg s$^{-1}$). The illuminated regions of a typical warped disk are shown in Figure 1 of~\citet{hic05}. In this scenario, the hard X-ray pulsar beam sweeps across and irradiates the inner disk, thereby producing soft X-ray emission. As the disk precesses, the phase difference between the soft and hard pulses is observed to vary with the superorbital phase. ~\citet{hic05} developed a simplified geometric model of a warped disk that successfully reproduced the long-term variations in the soft pulse profiles of SMC X-1, showing good agreement with observations. Using data from different superorbital phases of LMC X-4, ~\citet{hun10} reported variations in the phase shift between the hard and soft X-ray bands, which they interpreted as reprocessed hard X-rays from a precessing warped disk. More recently, ~\citet{bru20} analyzed data spanning a single superorbital cycle of both LMC X-4 and SMC X-1 and showed that relative changes between the hard and soft pulse profiles arise from reprocessed emission in a warped, precessing inner disk. This interpretation is further supported by the consistency between observed soft pulse profiles and those simulated at different superorbital phases using the warped-disk model of~\citet{hic05}.

In this study, we analyze 33 years of X-ray monitoring data from multiple instruments to investigate the superorbital modulation of LMC X-4. Section~\ref{obs} introduces the X-ray monitoring and scanning telescopes used to obtain the light curves, including CGRO BATSE, RXTE ASM, Swift BAT, MAXI GSC, and Fermi GBM. Data reduction procedures, including segmentation of light curves, folding, modeling of modulation profiles, and selection of fiducial points, are described in Section~\ref{smp}. The modeling of the long-term superorbital phase evolution trend using polynomial and sinusoidal functions, as well as the analysis of systematic phase fluctuations to estimate period variations over the 33-year baseline, are presented in Section~\ref{spe}. The discovery of a superorbital phase shift between soft and hard X-ray bands after MJD 57000 is reported in Section~\ref{ps}. Finally, Section~\ref{dis} discusses the superorbital period and phase evolution, including both the long-term trend and systematic fluctuations (Section~\ref{spv}), and interprets the soft-hard phase shift as a consequence of geometric changes in the warped disk (Section~\ref{sxps}).


\section{Observations} \label{obs}

\subsection{Burst and Transient Source Experiment on Compton Gamma Ray Observatory}
The Burst and Transient Source Experiment (BATSE) was one of four instruments aboard the Compton Gamma Ray Observatory (CGRO;~\citealt{geh93}), which operated from 1991 to 2000. BATSE was specifically designed to detect and characterize gamma-ray bursts (GRBs) and other transient high-energy phenomena. The instrument consisted of eight detector modules, each equipped with a Large Area Detector (LAD) and a spectroscopy detector. The LADs, constructed from 50.8 cm diameter NaI(Tl) scintillators, were optimized for high sensitivity in the 20 keV-1 MeV energy range ~\citep{fis94}. These modules were strategically positioned at the corners of the CGRO spacecraft to achieve near-complete sky coverage. BATSE's all-sky monitoring capabilities led to the pivotal discovery that GRBs are distributed isotropically, providing compelling evidence for their cosmological origin~\citep{mee92}. Beyond GRBs, BATSE yielded invaluable data on solar flares, pulsars, accreting X-ray binaries, and terrestrial gamma-ray flashes. In this study, we analyzed the daily-binned light curve of LMC X-4 in the 20-40 keV energy band, obtained from the website of the Gamma-Ray Astrophysics Team at the National Space Science and Technology Center (NSSTC)\footnote{\url{https://gammaray.nsstc.nasa.gov/batse/occultation/sourcepages/lmcx-4.html}}
. The dataset encompasses observations from May 30, 1991 (MJD 48406) to May 26, 2000 (MJD 51690).

\subsection{All-Sky Monitor on Rossi X-Ray Timing Explorer}
 The All-Sky Monitor (ASM;~\citealt{lev96}) was an instrument aboard Rossi X-Ray Timing Explorer (RXTE) designed to monitor variable and transient X-ray sources. The ASM consisted of three scanning shadow cameras, each equipped with a position-sensitive proportional counter, and used a one-dimensional coded mask to achieve a field of view of $6^{\circ} \times 90^{\circ}$. The ASM was sensitive to cosmic X-rays in the 1.5-12 keV range, which was subdivided into three sub-bands: 1.5-3 keV, 3-5 keV, and 5-12 keV. Light curves were produced with two temporal resolutions: individual 90-second ¡§dwell¡¨ exposures and one-day averages. Throughout its mission, from early 1996 to early 2012, the ASM scanned nearly the entire sky every 90 minutes. In this study, we used the 1.5-12 keV ASM light curve of LMC X-4 obtained from the HEASARC archive. Although the RXTE ASM remained operational until early 2012, data from the final years were compromised by elevated high-background events, likely attributable to Al K$_{\alpha}$ photons~\citep{lev11}. Therefore, following the selection criteria established by \citet{lev11}, we used data collected from January 5, 1996 (MJD 50087) to January 28, 2010 (MJD 55224) to investigate the superorbital modulation of LMC X-4.

\subsection{Burst Alert Telescope on Swift}
The Neil Gehrels Burst Alert Telescope (BAT) aboard the Swift Observatory is a coded-mask instrument with a wide field of view (1.4 steradians), optimized for monitoring the hard X-ray sky in the 15-150 keV energy band since 2004 \citep{bar05}. While the principal objective of BAT is the detection and localization of gamma-ray bursts, its substantial collecting area (5200 cm$^2$) and moderate angular resolution ($20\arcmin$) make it highly effective for continuous surveillance of known X-ray sources as the Swift completes an orbit around the Earth approximately every 96 minutes. This capability facilitates systematic investigations of long-term variability in these sources. In this study, we analyzed the daily-binned 15-50 keV Transient Monitor light curve~\citep{kri13} of LMC X-4, archived on the Swift website\footnote{\url{https://swift.gsfc.nasa.gov/results/transients/LMCX-4/}}, encompassing the period from February 15, 2005 (MJD 53416) to May 31, 2024 (MJD 60461).

\subsection{Gas Slit Camera on  Monitor of All-sky X-ray Image}
The Monitor of All-sky X-ray Image (MAXI), mounted on the Japanese Experiment Module of the International Space Station (ISS), is designed to detect transient X-ray phenomena and monitor the long-term variability of X-ray sources \citep{mat09}. MAXI comprises two types of slit cameras with different detectors: a Gas Slit Camera (GSC) with an effective area of 5250 cm$^2$, sensitive to the 2-30 keV range, and a Solid-state Slit Camera (SSC) with an effective area of 200 cm$^2$ covering the 0.5-12 keV band. MAXI achieves nearly complete sky coverage twice per ISS orbit, approximately every 90 minutes. In this study, we used the daily-binned light curve in the 2-10 keV energy band collected by the GSC and retrieved from the MAXI on-demand processing website\footnote{\url{http://maxi.riken.jp/mxondem}}. This energy range was selected to represent the soft X-ray emission and to ensure a clear spectral separation from the hard X-ray band (15-50 keV) observed by Swift BAT, thereby minimizing energy overlap in our comparison of the two bands. The dataset spans from August 12, 2009 (MJD 55055) to May 31, 2024 (MJD 60461) and was used to investigate the superorbital modulation of LMC X-4.

\subsection{Gamma-ray Burst Monitor on Fermi}
The Fermi Gamma-ray Burst Monitor (GBM) is an all-sky monitoring instrument aboard the Fermi Gamma-ray Space Telescope, launched in 2008 \citep{mee09}. Designed to detect gamma-ray bursts and other transient phenomena, it utilizes the Earth occultation technique \citep{wil12} and covers a broad energy range from 8 keV to 40 MeV. The GBM comprises 12 sodium iodide (NaI) detectors optimized for lower energies and two bismuth germanate (BGO) detectors for higher energies, strategically arranged to achieve nearly full-sky coverage. This configuration enables the rapid localization and characterization of transient gamma-ray phenomena, thereby facilitating follow-up observations by other instruments and ground-based facilities. In this study, we analyzed the daily-binned light curve of LMC X-4 in the 12-50 keV band, as provided by the NSSTC website\footnote{\url{https://gammaray.nsstc.nasa.gov/gbm/science/earth_occ/LMCX-4.html}}
. The dataset spans from February 13, 2012 (MJD 55970), the earliest available data, to May 31, 2024 (MJD 60461).\\

\begin{figure}
   \epsscale{1.2}
\plotone{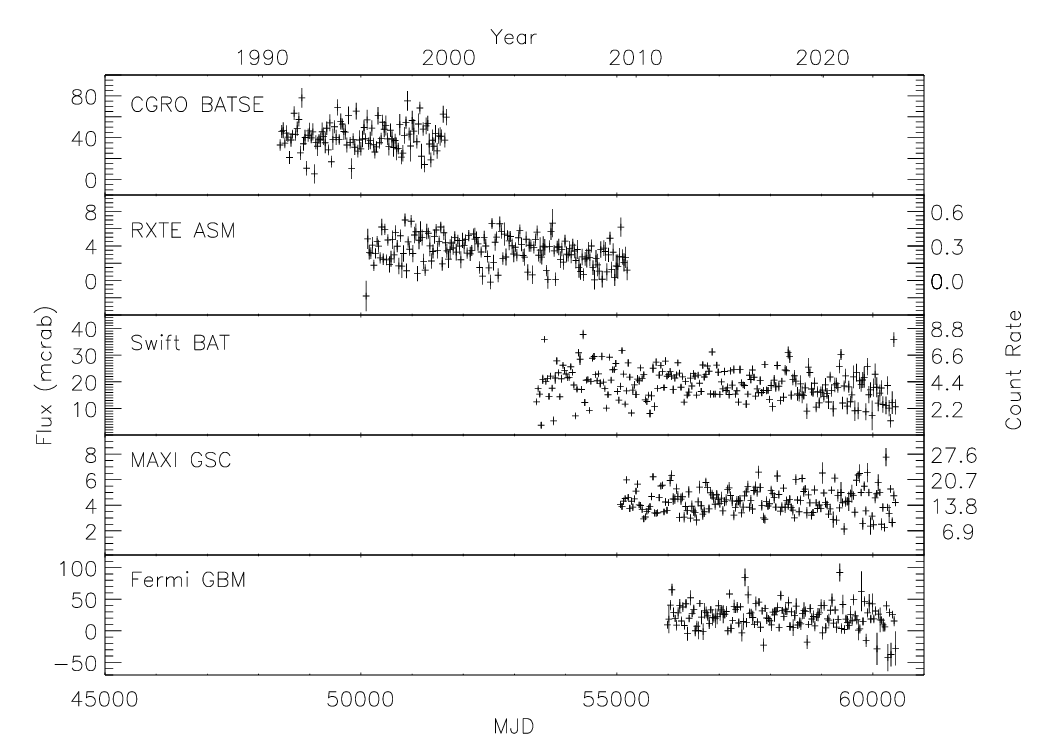}
\caption{Light curves collected by five instruments for analysis in this study. The bin size of these light curves is 30.34 days (about a superorbital cycle). The count rate units are cts s$^{-1}$ for RXTE ASM, $10^{-3}$ cts cm$^{-2}$ s$^{-1}$ for Swift BAT and $10^{-3}$ ph cm$^{-2}$ s$^{-1}$ for MAXI GSC. \label{lcurves}}
\end{figure}

The light curves of LMC X-4 collected by these five instruments are shown in Figure~\ref{lcurves} where 1 Crab corresponds to 75 cts s$^{-1}$ for RXTE ASM\footnote{\url{http://xte.mit.edu/asmlc/ASM.html}}, 0.22 cts cm$^{-2}$ s$^{-1}$ for Swift BAT\footnote{\url{https://swift.gsfc.nasa.gov/results/transients/}}, and 3.45 ph cm$^{-2}$ s$^{-1}$ for MAXI GSC\footnote{\url{http://maxi.riken.jp/top/readme.html}}.

\section{Data Analysis} \label{da}
\subsection{Superorbital Modulation Profile and Phase}\label{smp}

In this study, we investigate the evolution of the superorbital phase of LMC X-4. For phase analysis, it is standard practice to fold all light curves using a fixed linear ephemeris. To minimize ambiguities in cycle count caused by large phase drifts, we selected a folding period close to the mean period of the dataset. In this work, we adopted the optimal period for the RXTE ASM light curve, $P_f = 30.3187$ days, determined using the epoch folding search technique (i.e., {\it efsearch} in {\it Heasoft}), as the folding period. The maximum intensity was chosen as the fiducial point, consistent with the approach of \citet{hu15,mol15}. To determine the phase-zero epoch for folding  ($T_f$), we folded the RXTE ASM light curve using the folding period $P_f$ and an initial phase-zero epoch MJD 50085, the date of beginning of the RXTE ASM light curve. This revealed a significant phase shift of 0.38 for the fiducial point in the folded light curve. Accordingly, we adjusted the phase-zero epoch to $T_0 = \mathrm{MJD}$ 50075.472 to align the fiducial point with phase zero. This linear ephemeris

\begin{equation}\label{feph}
T_N = \mathrm{MJD} 50075.472 + 30.3187 N
\end{equation}

\noindent was employed for subsequent analysis. Although this ephemeris was derived from the RXTE ASM light curve rather than the full dataset, the superorbital modulation period of LMC X-4 is known to be stable, and thus no substantial phase drift is expected across the entire dataset. Under this ephemeris, the phase distribution ranges from -0.1 to 0.4 (see Section~\ref{spe}).

\begin{figure}
  \epsscale{1.4}
\plotone{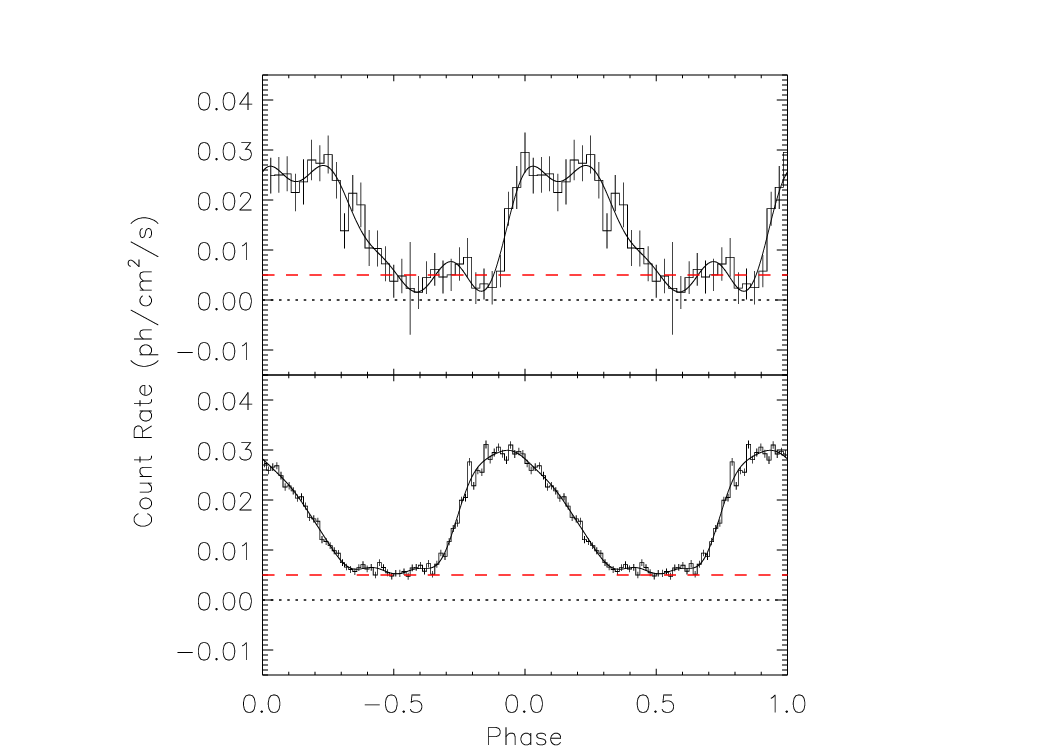}
\caption{Top: A typical superorbital modulation profile of a data segment, generated from the MAXI GSC light curve obtained between MJD 56351.44 and 56472.71, folded by the linear ephemeris in Eq.\ref{feph}. The dashed red line represents the baseline level for MAXI GSC (0.005 ph cm$^{-2}$ s$^{-1}$) and the dot black line represents the zero count rate. Bottom: Modulation profile obtained by folding the entire MAXI GSC light curve with the optimal sinusoidal ephemeris (Eq.~\ref{sinueph}).\label{expprof}}
\end{figure}

To investigate the evolution of the superorbital phase, we divided the light curves into segments and folded each segment using Eq.~\ref{feph}. For instruments with higher sensitivity to the superorbital modulation of LMC X-4, such as RXTE ASM, Swift BAT, and MAXI GSC, a clear superorbital profile can be obtained with a segment length of four cycles. Conversely, the same segment length was insufficient to yield a significant modulation profile for the CGRO BATSE and Fermi GBM light curves, necessitating a segment length of up to 20 cycles for these instruments. The selection of a fiducial point to define the superorbital phase has varied in previous studies. Some studies adopted the onset of the high state of the superorbital cycle (or its equivalent) as the fiducial point \citep[e.g.,][]{pau02,hun10,bru20}, although a precise definition of the high state was not provided. Alternatively, \citet{hu15} and \citet{mol15} fitted the modulation profile with a Gaussian function, taking its centroid as the fiducial point. However, our analysis reveals that the modulation profile is inherently asymmetric, as shown by the segmented profiles (top panel of Figure~\ref{expprof}) and the profile derived from the entire light curve (bottom panel of Figure~\ref{expprof}). Moreover, double-peaked structures occasionally appear in the segmented profiles (see top panel of Figure~\ref{expprof}). These results indicate that a Gaussian function does not adequately describe the superorbital modulation profile of LMC X-4.

Alternatively, we modeled the modulation profile using a multi-component sinusoidal function:

\begin{equation}\label{multisinu}
r(\phi) = a_0 + \sum_{k=1}^{n} \left[a_k \cos(2\pi k \phi) + b_k \sin(2\pi k \phi)\right]
\end{equation}

\noindent where $r$ represents the count rate, $\phi$ is the phase, and $n$ is the number of sinusoidal components. We found that all segmented profiles could be well described by a four-component sinusoidal function ($n=4$). To determine the phase of each profile, we adopted a method similar to that proposed by \citet{hu08}, using the centroid of the modulation as the fiducial point, defined as:

\begin{equation}\label{phase}
\phi_s = \frac{\sum_{i=1}^N (r_i - r_0)\phi_i}{\sum_{i=1}^N (r_i - r_0)}
\end{equation}

\noindent where $\phi_s$ is the fiducial point phase of the data segment, $r_i$ represents the fitted count rate from the four-component sinusoidal model, $\phi_i$ is the corresponding phase, and $r_0$ represents the baseline level of the profile. Although the baseline $r_0$ was difficult to identify in the segmented folded light curve, it was clearly visible in the profile constructed from the entire light curve (see bottom panel of Figure~\ref{expprof}). Based on the profiles from five instruments, we found that the baseline was approximately zero for RXTE ASM and Swift BAT, while it was approximately 0.005 cts cm$^{-2}$ s$^{-1}$, 5 mCrab, and 10 mCrab for MAXI GSC, CGRO BATSE, and Fermi GBM, respectively. We adopted $N=10^4$ a cycle, and the uncertainty of $\phi_s$ was estimated using $10^4$ runs of Monte Carlo simulations.

\subsection{Superorbital Phase Evolution}\label{spe}
\subsubsection{Polynomial Models}\label{pm}

\begin{figure}[htbp]
  \centering
  \includegraphics[width=0.45\textwidth]{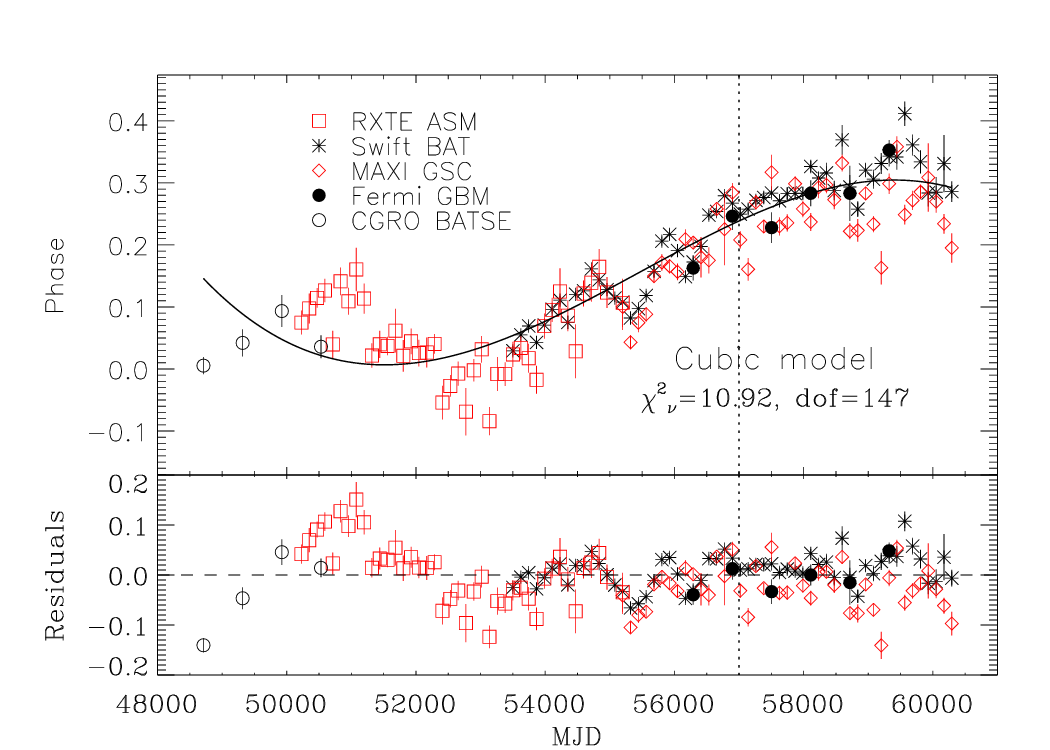}
   \includegraphics[width=0.45\textwidth]{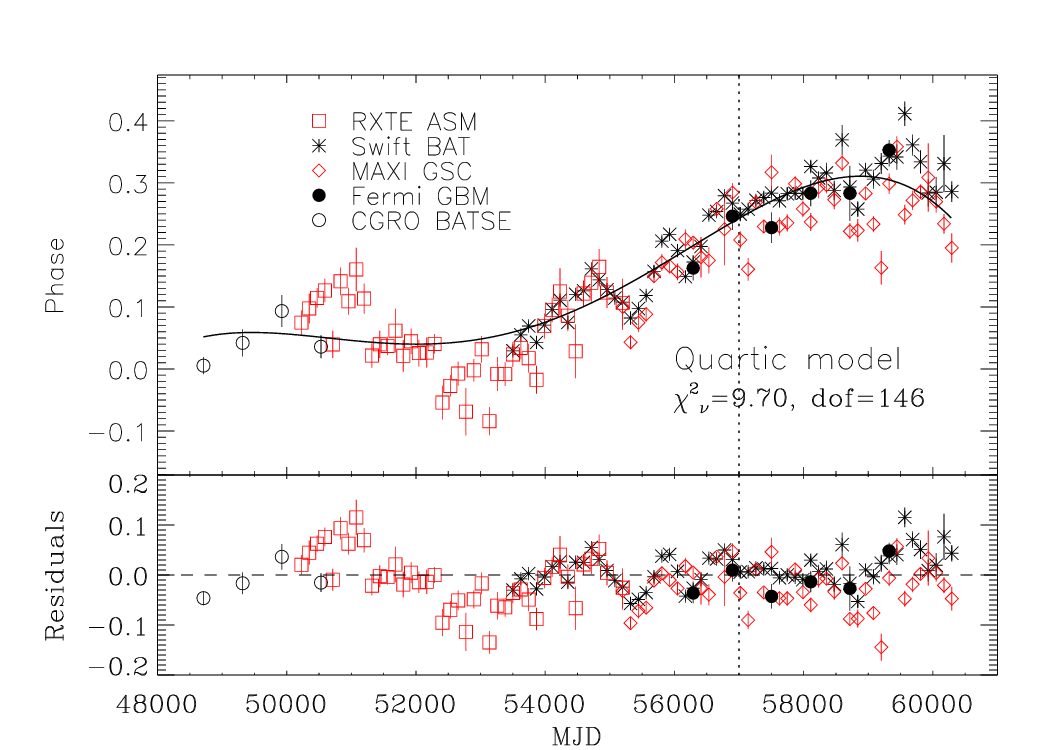}
 \includegraphics[width=0.45\textwidth]{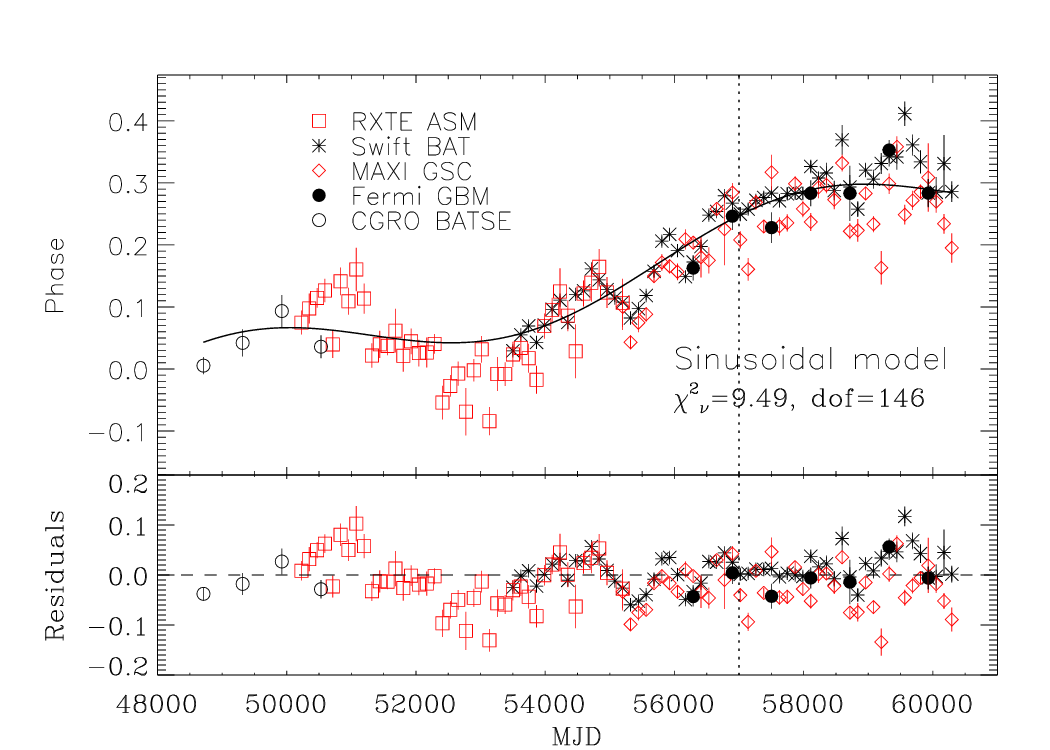}
\caption{Evolution of superorbital phases fitted by the cubic (top), quartic (middle), and sinusoidal (bottom) models from 1991 to 2024. Soft X-ray phases from RXTE ASM and MAXI GSC are plotted in red, while hard X-ray phases from CGRO BATSE, Swift BAT, and Fermi GBM are plotted in black. The solid lines represent the corresponding best-fit models describing the long-term phase evolution trend. The residuals obtained after subtracting the corresponding trend are shown below. The vertical dotted lines mark MJD 57000 after which phase shifts between soft and hard X-ray bands become apparent (Section~\ref{ps}).\label{phevplot}}
\end{figure}

\begin{figure}[htbp]
  \centering
  \includegraphics[width=0.45\textwidth]{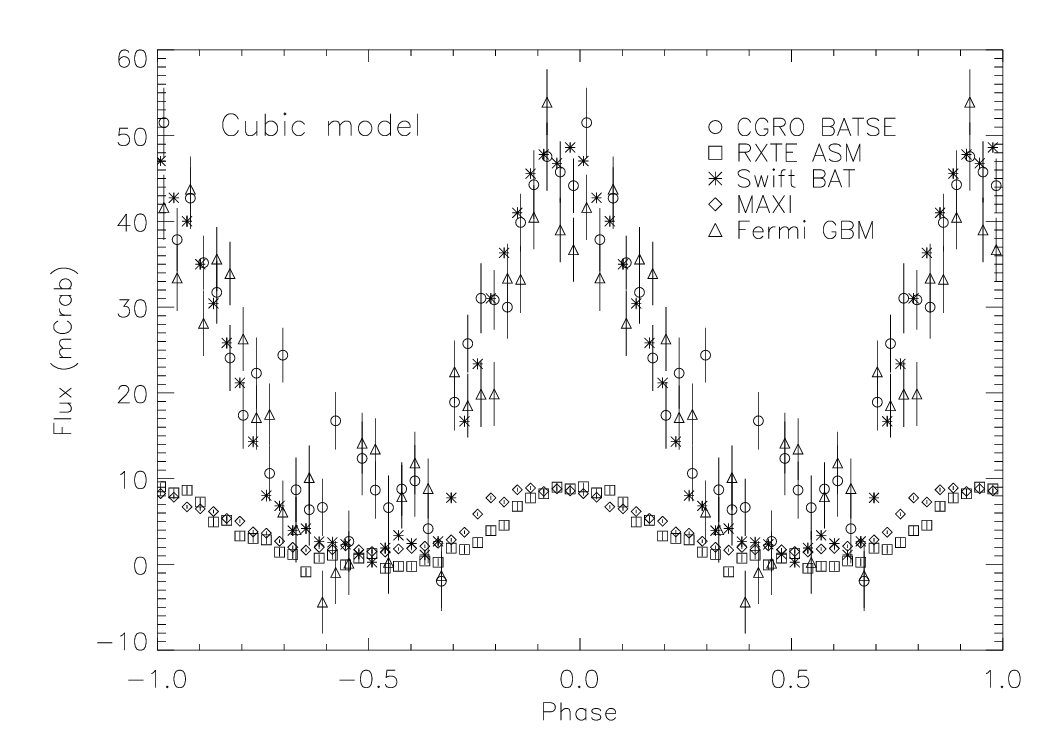}
   \includegraphics[width=0.45\textwidth]{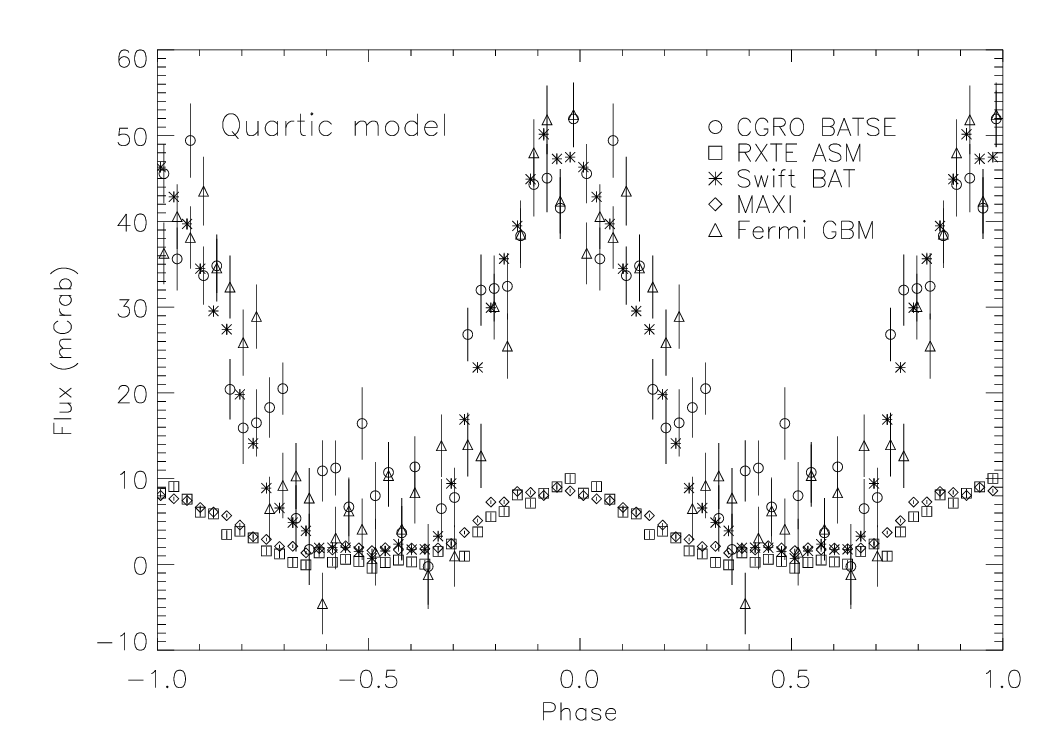}
 \includegraphics[width=0.45\textwidth]{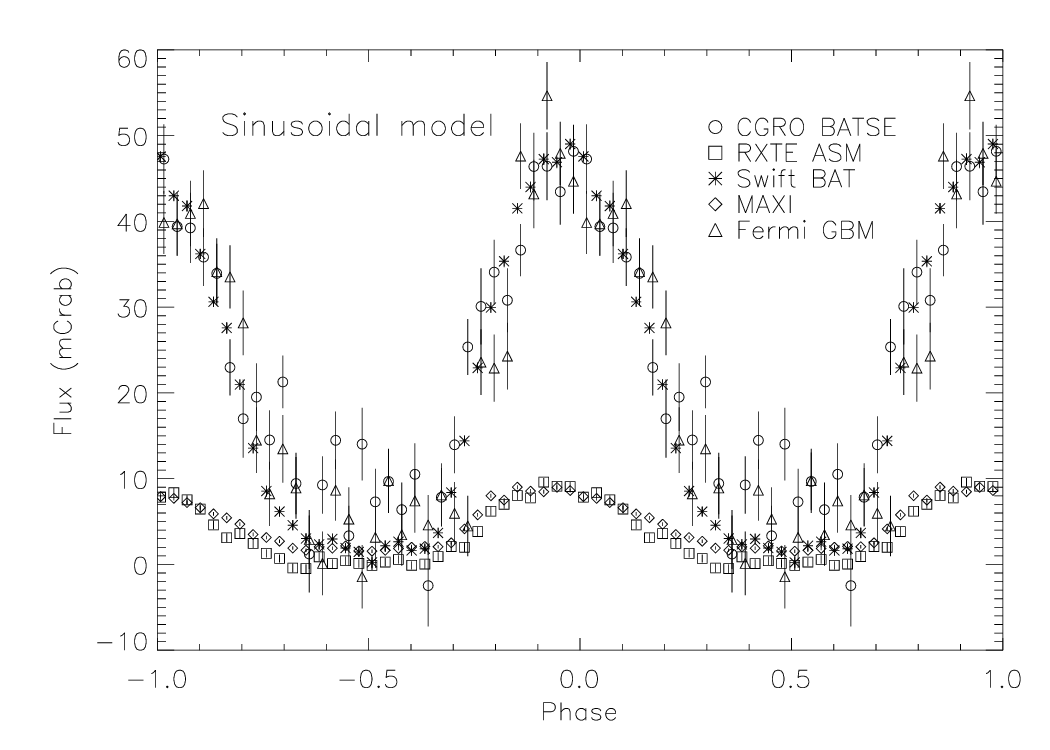}
\caption{Superorbital modulation profiles of the light curves from five instruments folded by the ephemerides derived from the cubic (top), quartic (middle), and sinusoidal (bottom) models.\label{flcplot}}
\end{figure}

The phase evolution of the superorbital modulation of LMC X-4, observed from May 30, 1991, to May 31, 2024, is shown in Figure~\ref{phevplot}. Consistent with the findings of~\citet{hu15}, the evolution demonstrated  a smooth long-term trend with superposed systematic fluctuations. However, visual inspection revealed that the quadratic function utilized by~\citet{hu15} failed to adequately describe the phase evolution trend. Therefore, we applied a cubic model to characterize the trend, with the best-fit parameters of the model listed in Table~\ref{para}. The phase evolution, cubic trend, and residuals are shown in the top panel of Figure~\ref{phevplot}, whereas the superorbital modulation profiles for the five instruments are shown in the top panel of Figure~\ref{flcplot}. The ephemeris of the cubic trend can be expressed as follows:

\begin{eqnarray}\label{ceph}
  T_N &=&   (MJD 50076.68 \pm 0.15)\\
\nonumber  & & +(30.2740 \pm 0.0028) \times N)\\
\nonumber  & & +(5.44 \pm 0.019) \times 10^{-4} \times N^2\\
\nonumber  & & +(-1.019\pm0.039) \times 10^{-6} \times N^3
\end{eqnarray}

\begin{table}
\begin{center}
  \caption{Superorbital Phase Evolution Parameters of LMC X-4\label{para}}

  \begin{tabular}{lll}
  \tableline\tableline
  \tableline\tableline
\multicolumn{3}{l}{Cubic model} \\
\multicolumn{3}{l}{$\phi=a_0 + a_1  (t-T_f) + a_2 (t-T_f)^2 + a_3 (t-T_f)^3$\tablenotemark{a}} \\
\multicolumn{3}{l}{$T_0 = T_f + a_0 P_f$\tablenotemark{b}} \\
\multicolumn{3}{l}{$P_0 = P_f /(1-a_1 P_f)$} \\
\multicolumn{3}{l}{${\dot P}_0 = 2 a_2 {P_0}^2$} \\
\multicolumn{3}{l}{$\ddot P = 6a_3 {P_0}^2$} \\
\tableline
Parameter& & Value \\
$a_0$ (cycle) &  &$0.0399 \pm 0.0049$\\ 
$a_1 $ (cycle/day) & &$(-4.87 \pm 0.30) \times 10^{-5}$ \\ 
$a_2 $ (cycle/day$^2$) & &$ (1.958 \pm 0.067) \times 10^{-8}$ \\
$a_3 $ (cycle/day$^3$) & &$(-1.213 \pm 0.045) \times 10^{-12}$ \\
$T_0$ (MJD) & & $50076.68 \pm 0.15$\\
$P_0$ (days)& &$0.2740 \pm 0.0028$\\
${\dot P}_0$ (day/day)& &$ (3.59 \pm 0.12) \times 10^{-5}$\\
$\ddot P$ (day/day$^2$)& &$(-6.707 \pm 0.025) \times 10^{-9}$\\
\tableline\tableline
\tableline\tableline
\multicolumn{3}{l}{Quartic model} \\
\multicolumn{3}{l}{$\phi=a_0 + a_1  (t-T_f) + a_2 (t-T_f)^2 + a_3 (t-T_f)^3+ a_4 (t-T_f)^4$\tablenotemark{a}}\\
\multicolumn{3}{l}{$T_0 = T_f + a_0 P_f$\tablenotemark{b}} \\
\multicolumn{3}{l}{$P_0 = P_f /(1-a_1 P_f)$} \\
\multicolumn{3}{l}{${\dot P}_0 = 2 a_2 {P_0}^2$} \\
\multicolumn{3}{l}{$\ddot P = 6a_3 {P_0}^2$} \\
\multicolumn{3}{l}{$\dddot P = 24a_4 {P_0}^2$} \\

\tableline
Parameter& & Value \\
$a_0$ (cycle) &  &$0.0558 \pm 0.0051$\\ 
$a_1 $ (cycle/day) & &$ (-8.8 \pm 4.2) \times 10^{-5}$ \\ 
$a_2 $ (cycle/day$^2$) & &$(-4.3 \pm 1.9) \times 10^{-9}$ \\
$a_3 $ (cycle/day$^3$) & &$(2.82 \pm 0.30) \times 10^{-12}$ \\
$a_4 $ (cycle/day$^4$) & &$(-2.10 \pm 0.15) \times 10^{-16}$ \\
$T_0$ (MJD) & &$50077.16 \pm 0.15$\\
$P_0$ (days)& &$ 30.3106 \pm 0.0039$\\
${\dot P}_0$ (day/day)& & $(-8.0 \pm 3.4) \times 10^{-6}$\\
${\ddot P}_0$ (day/day$^2$)& &$(1.56 \pm 0.16) \times 10^{-8}$\\
$\dddot P$ (day/day$^3$)& &$(-4.63 \pm 0.34) \times 10^{-12}$\\
\tableline\tableline
\tableline\tableline
\multicolumn{3}{l}{Sinusoidal model} \\
\multicolumn{3}{l}{$\phi=a_0 + a_1  (t-T_f) + a_2 cos[{{2\pi} \over a_4}(t-T_f)]+ a_3 sin[{{2\pi} \over a_4}(t-T_f)]$\tablenotemark{a}}\\
\multicolumn{3}{l}{$T_0 = T_f + a_0 P_f$\tablenotemark{b}} \\
\multicolumn{3}{l}{$P_0 = P_f /(1-a_1 P_f)$} \\
\multicolumn{3}{l}{$P_\mathrm{mod}=a_4$} \\
\tableline
Parameter&  & Value \\
$a_0$ (cycle)&  &$0.023 \pm 0.0059$\\ 
$a_1 $ (cycle/day) & &$(2.60 \pm 0.11) \times 10^{-5}$ \\ 
$a_2 $ (cycle) & &$0.0440 \pm 0.0035$ \\
$a_3 $ (cycle) & &$ -0.00376 \pm 0.0070$ \\
$a_4 $ (day) & &$ 8900 ^{+210}_{-230}$ \\
$T_0$ (MJD) & &$50076.16 \pm 0.18$\\
$P_0$ (days)& &$30.34263 \pm 0.00098$\\
$P_\mathrm{mod}$ (days)& &$ 8900 ^{+210}_{-230}$\\

\tableline\tableline\tableline

\end{tabular}
\tablenotetext{}{\parbox[t]{0.95\columnwidth}{\raggedright  Notes:}}
  \tablenotetext{a}{\parbox[t]{0.90\columnwidth}{\raggedright $T_f$=MJD 50075.472}}
  \tablenotetext{b}{\parbox[t]{0.90\columnwidth} {\raggedright $P_f$=30.3187 days}} 
\end{center}
\end{table}

The superorbital period of LMC X-4 is regarded as one of the most stable among all X-ray binaries whose superorbital modulation is driven by a warped, precessing accretion disk~\citep{cla03,ogi01,kot12}. In this study, the superorbital phase evolution suggests that  some superorbital period changes remain in the 33-year dataset. Excluding short-term systematic phase fluctuations, the maximum and minimum periods estimated from the cubic model parameters listed in Table~\ref{para} indicate a total period variation of 0.152 days, which is approximately 0.50\% of the superorbital period, validating that the superorbital period of LMC X-4 remains remarkably stable.

However, the phase evolution during the early observations, primarily from CGRO/BATSE between MJD $\sim$48500 and $\sim$50000, significantly deviates from the cubic model prediction. Although these discrepancies may be attributed to systematic fluctuations, we investigated whether incorporating a higher-order term would improve the fit by applying a quartic model. The cubic fit yields $\chi^2 = 1604.44$ with 150 degrees of freedom, whereas the quartic fit results in $\chi^2 = 1416.67$ with 149 degrees of freedom. An F-test produced an F-value of 19.77 and a corresponding p-value of $1.7 \times 10^{-5}$, demonstrating that the quartic model offers a statistically significant improvement. The quartic fit is shown in the middle panel of Figure~\ref{phevplot}, whereas the superorbital modulation profiles folded with the quartic ephemeris are shown in the middle panel of Figure~\ref{flcplot}. The best-fit parameters for the quartic model are listed in Table~\ref{para}, and the quartic ephemeris is expressed as follows:

\begin{eqnarray}\label{4eph}
  T_N &=& (MJD 50077.16 \pm 0.15)\\
  \nonumber   & &+(30.3106 \pm 0.0039) \times N\\
\nonumber  & &+(1.21 \pm0.52) \times 10^{-4}\times N^2\\
  \nonumber  & &+(2.39 \pm 0.25) \times 10^{-6} \times N^3\\
\nonumber  & &+(-5.36 \pm 0.39) \times 10^{-9} \times N^4
\end{eqnarray}

\noindent The maximum period change estimated from the quartic ephemeris was 0.156 days, approximately 0.52\% of the superorbital period. This result further supports the long-term stability of the superorbital period of LMC X-4.
Moreover, we evaluated a quintic model; however, the improvement in fit was marginal, with a p-value of 1.2\% from the F-test, and the derived parameters, including the period change, were consistent with those obtained from the quartic model. Given the empirical nature of polynomial models, we did not pursue higher-order fits in this analysis.

\subsubsection{Sinusoidal Model}\label{sm}

Alternatively, we explored a model comprising a linear trend with a superimposed sinusoidal modulation (hereafter, the sinusoidal model), as suggested by visual inspection of the data. The superorbital phase evolution was fitted using the following function:

\
\begin{eqnarray}\label{sinueph}
  \phi (t) &=& a_0 + a_1 (t - T_f) + a_2 \cos\left[ \frac{2\pi}{a_4} (t - T_f)\right]\\
\nonumber  & &+ a_3 \sin\left[ \frac{2\pi}{a_4} (t - T_f) \right]
\end{eqnarray}

\noindent where $T_f$ = MJD 50076.427 and $a_4 = P_{\mathrm{mod}}$ denotes the sinusoidal modulation period. The best-fit parameters are listed in Table~\ref{para}. This fit yielded a $\chi^2 = 1385.44$ with 149 degrees of freedom, comparable with that of the quartic model. According to this model, the superorbital phase, and consequently the superorbital period, undergoes modulation with a period of $P_{\mathrm{mod}} = 8900^{+210}_{-230}$ days.

To assess whether the inclusion of a quadratic term would improve the fit, we added this term to the model and performed an F-test. The resulting p-value was  51\%, indicating that the quadratic term is not statistically significant. Therefore, the ephemeris of the phase evolution trend based on the sinusoidal model, analogous to Equation (4) in~\citet{iar15}, can be expressed as follows:

\begin{eqnarray}\label{sinueph}
  T_N &=&(\mathrm{MJD}~50076.16 \pm 0.17 )\\
\nonumber & & + (30.34263 \pm 0.00098) \times N \\
\nonumber & & + (1.34 \pm 0.11) \times \cos\left[ \frac{2\pi}{N_{\mathrm{mod}}} N \right]\\
\nonumber & &+ (-0.14 \pm 0.21) \times \sin\left[ \frac{2\pi}{N_{\mathrm{mod}}} N \right]
\end{eqnarray}
\noindent where $N_{\mathrm{mod}} = P_{\mathrm{mod}} / P_0 =  293.3^{+7.6}_{-6.9}$. The sinusoidal fit is shown in the bottom panel of Figure~\ref{phevplot}, whereas the superorbital modulation profiles folded with the sinusoidal model ephemeris are shown in the bottom panel of Figure~\ref{flcplot}. The peak-to-peak period amplitude derived from this ephemeris corresponds to only 0.25\% of the mean superorbital period, further demonstrating the remarkable long-term stability of the superorbital modulation in LMC X-4.

\subsubsection{Systematic Phase Fluctuation and Superorbital Period Variation}\label{spv}

Similar to the results of~\citet{hu15}, the evolution of the superorbital phase is characterized by a smooth underlying trend with superimposed systematic phase fluctuations, as shown in Figure~\ref{phevplot}. Examination of the residuals reveals that these fluctuations, which remain within approximately ¡Ó0.1 cycles, are not random; however, they do not exhibit a clear pattern or periodicity. According to the empirical ephemeris proposed by~\citet{mol15} (Equation 2 and Table 2), the timescale of these systematic fluctuations is on the order of several hundred days, significantly shorter than the $\sim$9000-day variation timescale inferred from the overall phase evolution trend using the sinusoidal model.

Although the origin of these phase fluctuations remains unknown, we attempted to evaluate potential superorbital period variations from the phase residuals shown in Figure~\ref{phevplot}, assuming that these fluctuations result from changes in the superorbital period. Due to the phase shift between the soft and hard bands after MJD 57000 (see Section~\ref{ps}), which obscures the fluctuations, our analysis was restricted to data obtained prior to MJD 57000. According to the empirical ephemeris proposed by~\citet{mol15}, the superorbital period appeared to have experienced several ¡§glitches¡¨ within the timespan of their dataset. To investigate this behavior, we divided our phase residuals into several epochs using the time intervals ($T_1$ and $T_2$) defined in Table 2 by~\citet{mol15}. For each epoch, we fitted a linear function to the residual phases to estimate the variations in the superorbital period. The root-mean-square (rms) values of the period changes for each model are listed in Table~\ref{rmsp}. Notably, these rms values are significantly larger than those derived from the long-term phase evolution trends. Assuming that the superorbital period variations inferred from the systematic fluctuations and those from the long-term trend are independent, given their distinct timescales, the total rms period variation over this 33-year interval (MJD 48406 to 60461) is only about 0.55\%. This finding further underscores the remarkable stability of the superorbital modulation period of LMC X-4. A more detailed discussion of these short-timescale phase fluctuations is provided in Section~\ref{spv}.

\begin{table}
\begin{center}
  \caption{Rms of Superorbital Period Variation\label{rmsp}}

  \begin{tabular}{ccccccc}
    \\
    \tableline\tableline
     \tableline\tableline
     Model & $P_{mean}$ & $\sigma_{res}$\tablenotemark{a}& $\sigma_{trend}$ \tablenotemark{b}& $\sigma_{total}$ \tablenotemark{c}& $\sigma_{total}/P_{mean}$ \\
           & (days)  & (days)        & (days)       & (days)         & (\%) \\ 
\tableline
Cubic      & 30.330 & 0.152 & 0.040  & 0.157 & 0.52\\
Quartic    & 30.334 & 0.156 & 0.033  & 0.159 & 0.53 \\
Sinusoidal & 30.343 & 0.166 & 0.027  & 0.168 & 0.55 \\

\tableline\tableline
\tableline\tableline
  \end{tabular}
      \tablenotetext{}{\parbox[t]{0.95\columnwidth}{\raggedright Notes:}}
  \tablenotetext{a}{\parbox[t]{0.90\columnwidth} {\raggedright Rms of the period variation evaluated from the phase residuals of cubic, quartic and sinusoidal trends.}}
  \tablenotetext{b}{\parbox[t]{0.90\columnwidth} {\raggedright Rms of the period variation evaluated from the cubic, quartic and sinusoidal trend.}} 
  \tablenotetext{c}{\parbox[t]{0.90\columnwidth} {\raggedright Total rms superorbital variations with $\sigma_{total}=\sqrt{\sigma_{trend}^2 + \sigma_{res}^2}$ assuming that the superorbital period variations from systematic phase variations and long-term phase trend are independent.}}

\end{center}

\end{table}

\subsection{Phase Shift Between Soft and Hard X-ray Bands}\label{ps} 

By visually inspecting the superorbital phase evolution and the residuals shown in Figures~\ref{phevplot}, following the start of the Swift BAT observations (MJD 53416), we found that the hard X-ray band phases detected by Swift BAT and Fermi GBM, and the soft X-ray band phases detected by RXTE ASM and MAXI GSC, are generally well aligned prior to MJD $\sim$57000. However, beyond this epoch, the soft X-ray band phases appear to lead the hard X-ray band phases.

\begin{figure}[htbp]
  \centering
  \includegraphics[width=0.45\textwidth]{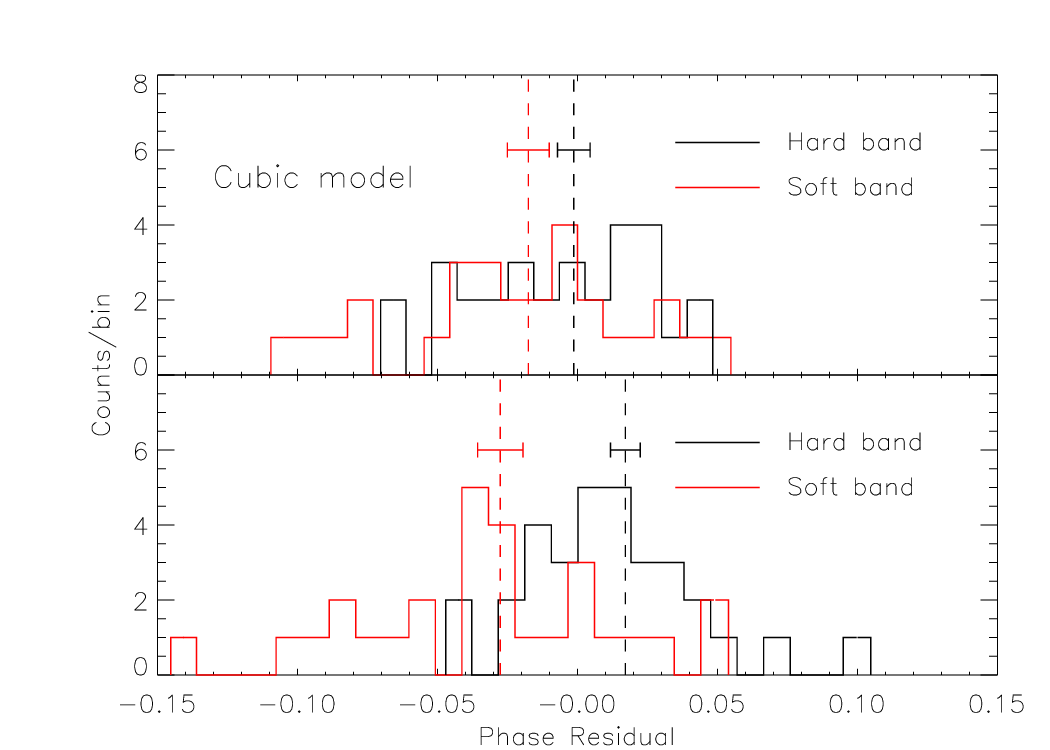}
   \includegraphics[width=0.45\textwidth]{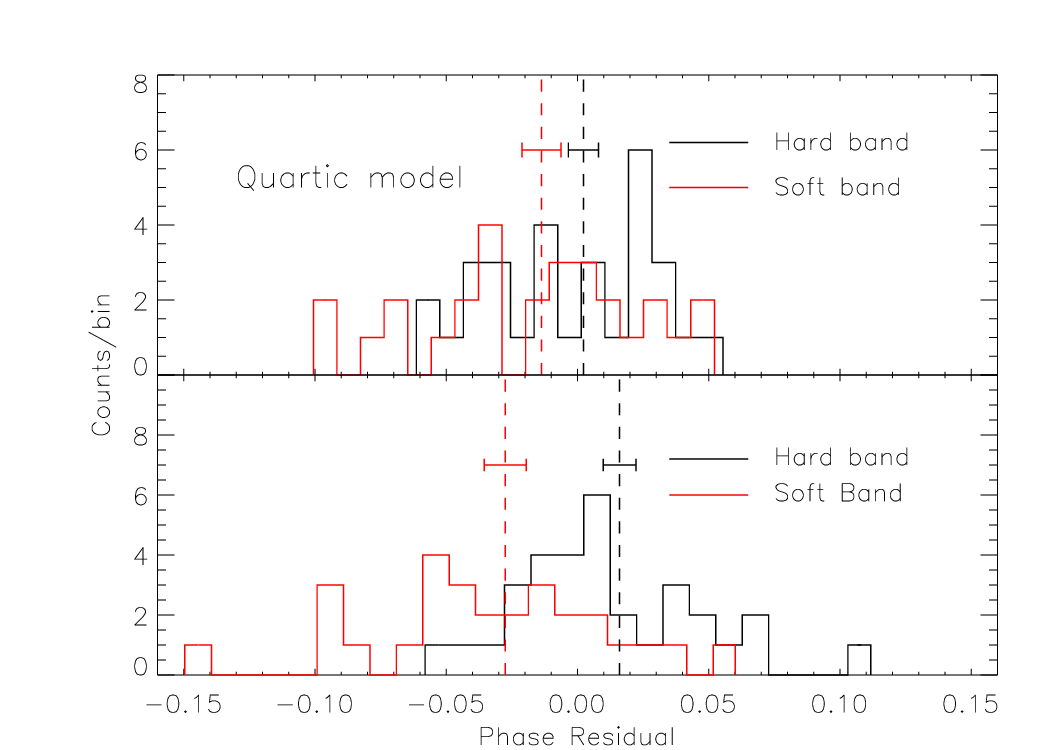}
 \includegraphics[width=0.45\textwidth]{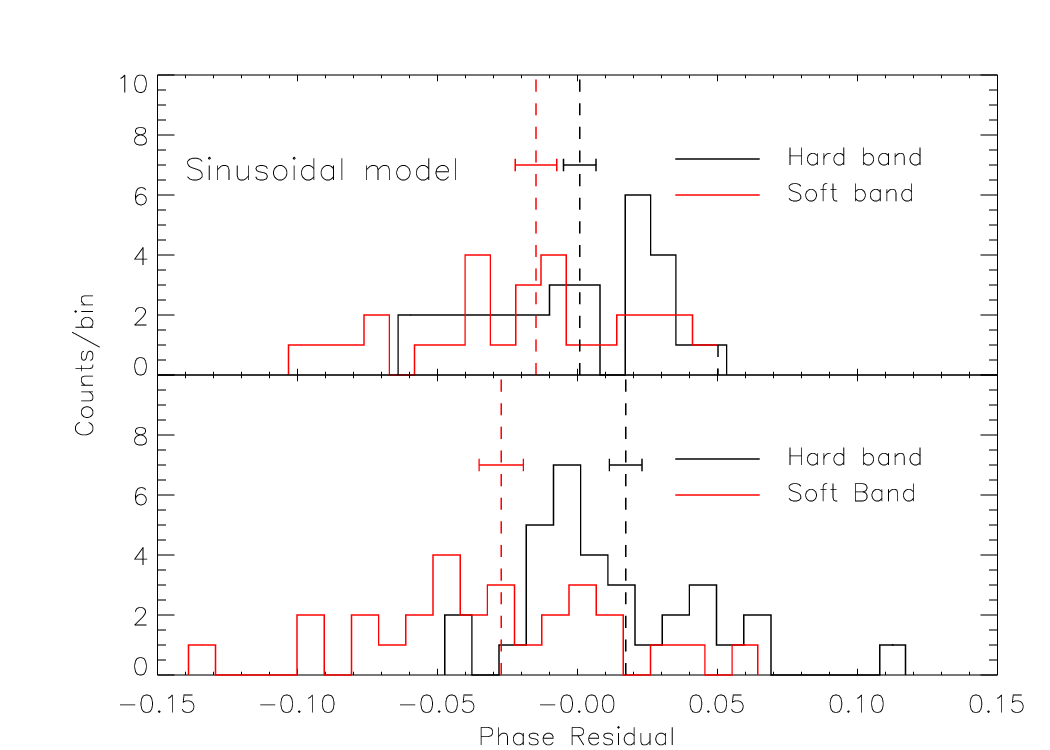}
\caption{Comparison of superorbital phase distributions between soft (red) and hard (black) X-ray bands between MJD 53500 to 57000 (Part 1, top panel) and MJD 57000 to 60300 (Part 2, bottom panel), where the phases are evaluated by utilizing the ephemerides derived from cubic (top), quartic (middle), and sinusoidal (bottom) models. The dashed lines and horizontal error bars indicate the corresponding mean values and their 1$\sigma$ uncertainties, respectively. \label{phdist}}
\end{figure}

To test for a potential phase shift, we divided the residuals into two intervals: Part 1 (MJD 53416-57000) and Part 2 (MJD 57000-60461). The distributions of soft- and hard-band phase residuals derived from the segmented light curves fitted with the cubic, quartic, and sinusoidal models are shown in Figure~\ref{phdist}. Noticeable differences between the distributions are presented in Parts 1 and 2. The corresponding mean phase differences (${\bar \phi}_{\mathrm{soft}} - {\bar \phi}_{\mathrm{hard}}$) are listed in Table~\ref{phaseshift} (Method 1). Using the cubic ephemeris as an example, the significance of the phase difference is 1.68$\sigma$ in Part 1 and 4.40$\sigma$ in Part 2. Additionally, a Student's $t$-test was performed to evaluate the statistical significance of these differences, yielding $p$-values of 9.75\% for Part 1 and 0.003\% for Part 2, as shown in Table~\ref{phaseshift}. These results indicate that the phase shift is not significant in Part 1 but becomes highly significant in Part 2. Similar results were obtained by the quartic and sinusoidal ephemerides, as listed in Table~\ref{phaseshift}.

\begin{table*}[ht]
\centering
  \caption{Soft X-ray Band Phase Shift Relatieve to Hard X-ray Band\label{phaseshift}}

  \begin{tabularx}{\textwidth}{cc|ccc|ccc}
    \\
    \tableline\tableline
     \tableline\tableline
  &   &  & Part1             &  &   & Part 2 & \\
 
\tableline
& & Method 1 \tablenotemark{a}& Method 2\tablenotemark{b} & Method 3\tablenotemark{c} &  Method 1 & Method 2 & Method 3\\
 \tableline
 Duration (MJD)& & 53416-57000 & 55055-57000 &  55055-57000 &  57000-60461 &57000-60461  & 57000-60461\\
  \tableline
  Phase shift (cubic ephemeris) & & -0.0163(97)& 0.001(15) & 0.0010(23) &-0.044(10) & -0.0292(62) & -0.0156(22)  \\
  Significance & & 1.68$\sigma$& 0.07$\sigma$ & 0.43$\sigma$ &4.40$\sigma$ &4.71$\sigma$ &  7.10$\sigma$ \\
  P-value from Studnet's t-test (\%) & &9.74 &-&- &0.003 & - & -  \\
  \tableline
Phase shift (quartic ephemeris) & & -0.0160(95) &-0.002(16) & 0.0028(23)  & -0.044(11) & -0.0191(62) & -0.0074(24) \\
Significance & &1.68$\sigma$&0.14$\sigma$ & 1.22$\sigma$ &4.00$\sigma$ &$3.08\sigma$ & 3.08 $\sigma$ \\
  P-value from Studnet's t-test (\%) & &9.75&-&- &0.010 & - & -  \\
  \tableline
Phase shift (sinusoidal ephemeris) & & -0.0156(95) & 0.0019(17) &-0.0029(24)&-0.044(10)&-0.0238(59) & -0.0105(23)       \\
Significance & & 1.64$\sigma$& 1.12$\sigma$ & 1.21$\sigma$ &4.40$\sigma$ &4.03$\sigma$ &4.57$\sigma$ \\
  P-value from Studnet's t-test (\%) & &10.75&-&- &0.005& - & -  \\
\tableline\tableline
\tableline\tableline
\end{tabularx}
  \tablenotetext{}{\raggedright }
  \tablenotetext{}{\raggedright  Notes:}
  \tablenotetext{a}{\raggedright Difference of the mean phases of soft and hard X-ray bands  (${\bar \phi}_{soft} - {\bar \phi}_{hard}$) from segmented light curves during the corresponding time duration.}
  \tablenotetext{b}{\raggedright Phase difference of the folded light curves collected by MAXI GSC and Swift BAT during the corresponding time duration.}
  \tablenotetext{c}{\raggedright Phase difference evlauaed by the cross correlation of the fitted modulation profiles collected by MAXI GSC and Swift BAT during the corresponding time duration.}

\end{table*}

Furthermore, we applied two supplementary methods to evaluate the significance of the phase shift using the overall modulation profiles folded with the relevant ephemerides (cubic, quartic, and sinusoidal) for Parts 1 and 2: (1) Measuring the phase difference of the centroids, that is, the fiducial points of the modulation profiles (Method 2). (2) Computing cross-correlation functions (Method 3). The modulation profiles used in Methods 2 and 3 were generated by folding each light curve (Part 1 or Part 2) into 64 phase bins using the corresponding ephemeris, and fitting the folded profiles with the n=6 harmonic model of Eq. (2) to enhance the profile structure. For Method 3, the phase lag was then obtained by cross-correlating the fitted soft- and hard-band modulation profiles, and the uncertainty was estimated by $10^4$ runs Monte-Carlo simulation. However, owing to differences in instrumental response, combining modulation profiles from different instruments (e.g., RXTE ASM and MAXI GSC in the soft X-ray band) remains challenging. Consequently, the comparisons were limited to MAXI GSC (soft X-ray band, 2-10 keV) and Swift BAT (hard X-ray band, 15-50 keV) for Method 2 and 3. Accordingly, the start time of Part 1 in these analyses was set to the beginning of the MAXI GSC observations (MJD 55055). The phase differences estimated using these two methods are listed in Table~\ref{phaseshift}. The significances of the phase differences are less than 2$\sigma$ for Part 1 but exceed 3$\sigma$ for Part 2, consistent with the results from Method 1, although the measured phase difference values are somewhat smaller. This discrepancy is likely attributable to smearing of the overall profiles caused by systematic phase fluctuations.

\begin{figure}
  \epsscale{1.2}
\plotone{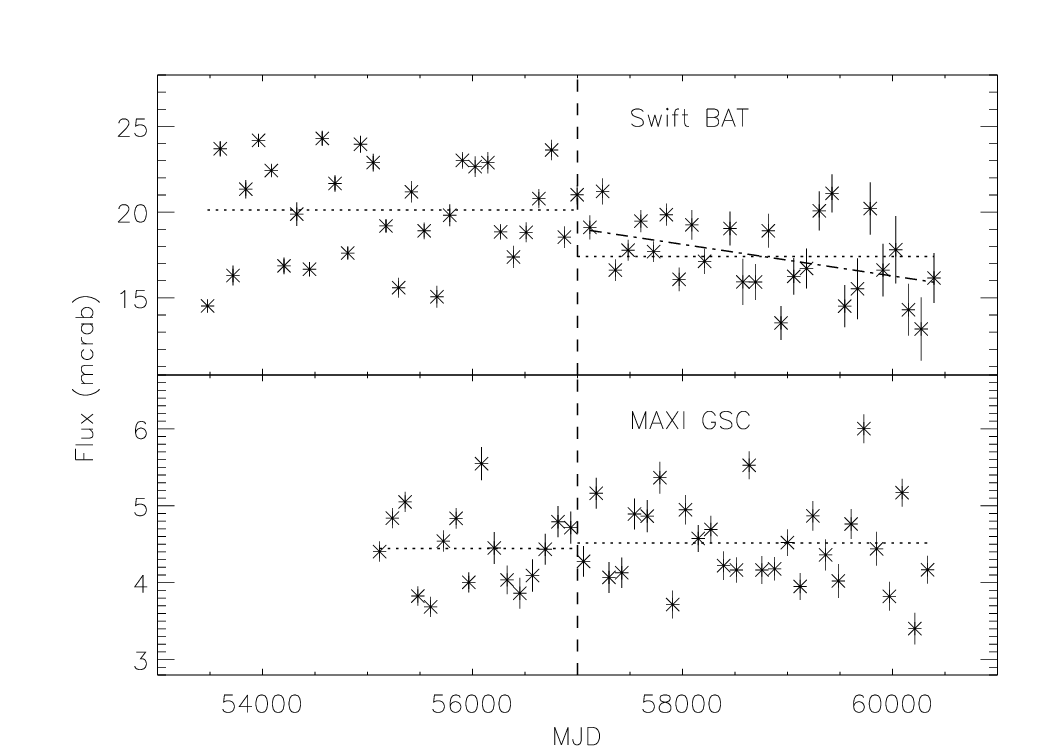}
\caption{Top: Swift BAT light curve with a bin size of 121.37 days (about 4 superorbital cycles). The vertical dashed line marks MJD 57000,whereas the dotted lines represent the mean fluxes before and after MJD 57000. Moreover, the dash-dotted line shows the decline in the flux after MJD 57000 with a mean decline rate of $(-8.4 \pm 2.0) \times 10^{-4}$ mCrab day$^{-1}$. Bottom: Similar to that of top panel but represents the MAXI GSC light curve. \label{swiftlc}}
\end{figure}

To investigate the origin of the observed phase shift, we examined the light curves obtained from Swift BAT and MAXI GSC. From the Swift BAT light curves, binned at 121.37 days (approximately four superorbital cycles; Figure~\ref{swiftlc}), we measured mean fluxes of $20.12 \pm 0.55$ mCrab and $17.41 \pm 0.44$ mCrab for the epochs before and after MJD 57000, respectively, corresponding to a significant decrease of approximately 13.5\%. This difference of $2.71 \pm 0.70$ mCrab corresponds to a 3.9$\sigma$ significance, and a Student's $t$-test yielded a $p$-value of 0.028\%. Moreover, as shown in Figure~\ref{swiftlc}, the epoch after MJD 57000 exhibits a declining trend, which likely contributes to the reduced mean flux. In contrast, the MAXI GSC light curve shows no significant change, with only a 0.41$\sigma$ flux difference and a $p$-value of 69\% from the Student's $t$-test.

In summary, a significant phase shift between the soft and hard X-ray bands is observed during MJD 57000-60461 but not during MJD 53416-57000. This phase shift coincides with the change in the hard X-ray flux. Further implications of the superorbital modulation phase shift between the soft and hard X-ray bands are discussed in Section~\ref{sxps}.

\section{Discussion} \label{dis}
\subsection{Superorbital Period Variations} \label{spv}

Superorbital modulations, characterized by periods much longer than the corresponding orbital periods, have been observed in numerous HMXBs and low-mass X-ray binaries (LMXBs). The superorbital period can be stable (e.g., Her X-1, LMC X-4, and SS 433), variable (e.g., SMC X-1), or even intermittent (e.g., Cyg X-2)~\citep{kot12}. Similar to Her X-1 and SMC X-1, the superorbital modulation of LMC X-4 has been attributed to periodic occultation of the central X-ray-emitting region, including the neutron star and the inner disk, by a precessing warped or tilted accretion disk~\citep{lan81,gru84,woj98,ogi01,cla03,hun10,kot12,bru20}. Superorbital behavior is also seen in Be/X-ray binaries~\citep[e.g., A0538-66;][]{alc01}, likely driven by a misaligned, precessing Be disk~\citep{mar24}. In wind-fed supergiant systems~\citep[e.g., 2S 0114+650;][]{cor99}, the superorbital modulations are possibly caused by interactions with corotating interaction regions around the donor star~\citep{boz17,cor21,isl23}. Alternatively, the tidal oscillation model has been proposed to drive these variations~\citep{koe06}, although its applicability may be limited in systems with eccentric orbits, as it generally necessitates a circular orbit. In LMXBs~\citep[e.g., 4U 1820-30;][]{pri84}, they may result from mass-loss rate variations induced by a third body~\citep{chou01} or from irradiation-driven instabilities on the companion~\citep{chou25}.

\citet{ogi01} demonstrated, through an analysis of radiation-driven disk-warping mechanisms in accretion disks, that the superorbital variations observed in LMC X-4, as well as in Her X-1 and SS 433, are situated near the stability limit for plausible system parameters. The stability of the superorbital periodicity of LMC X-4 was further supported by~\citet{cla03}, who applied dynamic power spectral analysis to the RXTE ASM light curve obtained from January 1996 to August 2002. A subsequent investigation by~\citet{kot12} showed that the superorbital period of LMC X-4 remained stable for approximately 15 years, based on a dynamic power spectral analysis of the RXTE ASM light curve spanning February 20, 1996, to February 12, 2011. However, in dynamic power spectral analysis, the frequency and time resolutions are constrained by $\Delta f \times \Delta T \sim 1$, where $\Delta f$ represents the peak width of the signal in the power spectrum and $\Delta T$ is the window size. Consequently, this method is less sensitive to small period variations or rapid, short-timescale changes in the period.

In contrast, phase analysis, equivalent to the observed-minus-calculated (O-C) method, enables a more precise determination of the period and other parameters (e.g., $\dot{P}$), provided that the modulation profile is stable and the fiducial point is well defined~\citep{chou14}. As shown in Figures~\ref{phevplot}, the superorbital phase evolution of LMC X-4 exhibits a smooth long-term trend superimposed with systematic phase fluctuations. To model this trend, we fitted the phase evolution using cubic, quartic, and sinusoidal functions. The quartic model provides a better fit than the cubic model and performs comparably with the sinusoidal model. Based on the superorbital periods derived from these ephemerides, the total period change over 33 years does not exceed 0.55\%, highlighting the remarkable stability of the superorbital modulation of LMC X-4.

The sinusoidal model suggests that the superorbital period of LMC X-4 varies periodically. Similar long-term variations have been reported in other X-ray binaries. For example, Doppler-shift analyses of SS 433 by~\citet{and83} and~\citet{mar89} showed that its 164-day disk-precession period varies with characteristic timescales of $1630 \pm 310$ days and $1299 \pm 24$ days. Quasi-periodic phase variations have also been detected in the 35-day superorbital modulation of Her X-1:~\citet{dav74} and \citet{boy80} found variations on timescales of a few hundred days, and~\citet{sta09,sta13} reported a $\sim$5-year variation, although the origin of this 5-year modulation remains unclear~\citep{sta09}. These cases provide context for interpreting the 8900-day superorbital variation observed in LMC X-4.

In the case of LMC X-4, our sinusoidal modeling indicates that the superorbital phase is modulated with a period of $P_{\mathrm{mod}} =  8900$ days and an amplitude of $A_{\phi} = 0.0578$ cycles. If this variation originates from the binary system orbiting a third companion, the projected semimajor axis would be $a_3 \sin i_3 = c A_{\phi} P_{\mathrm{sup}} = 1.574$ lt-day, where $c$ is the speed of light. The corresponding radial velocity amplitude would be $K = 2 \pi a_3 \sin i_3 / P_{\mathrm{mod}} = 372$ km/s. For the LMC X-4 binary system, assuming a neutron star mass $M_n = 1.57 M_{\odot}$ and a companion mass $M_c = 18 M_{\odot}$~\citep{fal15}, the mass function yields a lower limit for the third body's mass of $4.69 \times 10^4 M_{\odot}$, implying the presence of an intermediate-mass black hole. If the LMC X-4 binary system indeed orbits such an object, the orbital Doppler effect would cause variations in the observed pulsation period with an amplitude of $K P_{\mathrm{spin}}/c = 1.67 \times 10^{-2}$ s, where $P_{\mathrm{spin}} = 13.5\ \mathrm{s}$ is the pulsation period, and with a period of $P_{\mathrm{mod}} =  8900$ days.

A sinusoidal-like spin-period variation has been observed in the LMC X-4 system, with an amplitude of $5.3 \times 10^{-4} \ \mathrm{s}$ and a period of 5090 days~\citep{mol17}. However, these parameters are inconsistent with those derived from our phase analysis. Moreover, as discussed by~\citet{mol17}, even the observed sinusoidal-like spin-period variation is unlikely to result from an orbital Doppler effect caused by a third body. This conclusion is supported by radial velocity measurements of LMC X-4 from optical absorption and emission lines, which consistently yield values of approximately $300 \ \mathrm{km \ s^{-1}}$ across different epochs~\citep[see][and references therein]{mol17}. In addition, the presence of a massive third body would produce strong perturbations on the binary orbit that should be detectable in pulsar timing residuals \citep[e.g.,][]{ran14}. However, no such perturbations are observed in the pulsar timing residuals of LMC X-4, making this scenario implausible. Consequently, the observed sinusoidal trend in the superorbital phase evolution is unlikely to be attributable to an external third body. Instead, it is more plausibly explained by the presence of an additional clock within the system, likely arising from disk dynamics. In particular, non-linear radiation-driven warping may naturally generate long-timescale quasi-periodicities in the precession rate. Cyclic variations in mass transfer or resonances between the orbital and disk-warping timescales can also modulate the period. Although our data cannot uniquely distinguish among these scenarios, a disk-dynamical origin is more consistent with the modest amplitude and long timescale of the modulation.

The systematic phase fluctuations, as evidenced by the residuals shown in Figures~\ref{phevplot}, indicate an atypical phase evolution. While these fluctuations are not pure white noise, their autocorrelation function resembles a $\delta$-function. The data suggest that the superorbital modulation experienced several period glitches at unpredictable epochs, with time intervals ranging from approximately 100 to 1000 days, much shorter than the timescale of the smooth trend variations ($\sim$9000 days from the sinusoidal model). Therefore, we estimated the mean periods within the time intervals defined by~\citet{mol15} and found that the rms of period variations resulting from systematic phase fluctuations was substantially larger than that derived from the corresponding smooth trends (Table~\ref{rmsp}). Although these systematic phase fluctuations dominate the superorbital period variations, the total rms of the period variation was only about $ 0.55\%$, indicating the overall stability of the superorbital period of LMC X-4. In contrast,~\citet{cla03} reported period jitters of $\pm 0.05$ days ($\pm 0.16\%$) in their dynamic power spectrum of LMC X-4, which are likely manifestations of the same systematic phase fluctuations. The smaller value reported by~\citet{cla03} can be attributed to the smearing of period variations by the finite window function.

Such systematic phase fluctuations have also been observed in Her X-1. In the O-C plots, equivalent to the superorbital evolution shown in this study (i.e., Figure \ref{phevplot}).~\citet{sta09} (bottom panel of Figure 3) and~\citet{sta13} (bottom panels of Figures 2 and 5) demonstrated fluctuations in Her X-1 similar to those seen in LMC X-4. Although~\citet{sta09} noted that the primary timescale of these variations was approximately 5 years ($\sim$2000 days), smaller-scale variations on timescales of several hundred days are also evident, particularly in the interval MJD 40000¡V45000. The O-C fluctuations estimated by \citet{sta09} were about 3$P_{\rm orb}$ (5.1 days, corresponding to 0.15 superorbital cycles), which is slightly larger than those observed in LMC X-4. The physical origin of these unusual phase evolutions remains unknown.~\citet{sta13} suggested that additional torques acting on the warped disk could be responsible. Alternatively, these phase fluctuations may stem from variations in the fiducial point (e.g., the turn-on clock in Her X-1). Similar fiducial-point variations have been observed as timing noise in accreting millisecond pulsar systems~\citep{chou08,kul13,pat21}, and as phase jitters in dipping LMXBs when X-ray dips were used as fiducial points to probe orbital-period evolution \citep{chou01,hu08}. In both LMC X-4 and Her X-1, such fiducial-point variations may have been caused by changes in the disk's shape.

\subsection{Soft X-ray Phase Shift and Warped Disk Shape Change}\label{sxps}

In this study, we observed a small ($0.044 \pm 0.10$ cycles) but statistically significant ($>3\sigma$) superorbital phase shift between the soft and hard X-ray bands in the light curves obtained between MJD 57000 and 60461. No such shift was observed in the earlier data spanning MJD 53500-57000 (see Figure~\ref{phdist} and Table~\ref{phaseshift}). This phase shift likely resulted from a change in the geometry of the warped disk. In the warped disk model shown in Figure 1 of~\citet{hic05}, the hard X-ray emission is dominated by the pulsar beam, whereas the soft component originates from the reprocessing of this beam by the warped accretion disk~\citep{hun10,bru20,amb22}. We rewrite the warped disk equation introduced by~\citet{hic05} as

\begin{equation}\label{hvdisk}
\theta_{\mathrm{d}} (r,\phi) = \Bigl[ \theta_{\mathrm{in}} + (\theta_{\mathrm{out}} - \theta_{\mathrm{in}}) \frac{r-r_{\mathrm{in}}}{r_{\mathrm{out}} - r_{\mathrm{in}}} \Bigr] \sin(\phi-\phi_{\mathrm{tw}})
\end{equation}

\noindent Because the azimuthal function contains only a single sinusoidal term, we have $\theta_{\mathrm{d}}(r,\phi+\pi) = -\theta_{\mathrm{d}}(r,\phi)$, indicating that the disk is antisymmetric. According to the schematic of the warped disk shown in Figure 1 of~\citet{hic05}, the superorbital modulation profiles of the hard and soft X-ray bands are in phase as the disk precesses, which is consistent with the superorbital profiles derived from the light curves obtained between MJD 53500 and 57000.

\begin{figure}[htbp]
  \centering
  \includegraphics[width=0.4\textwidth]{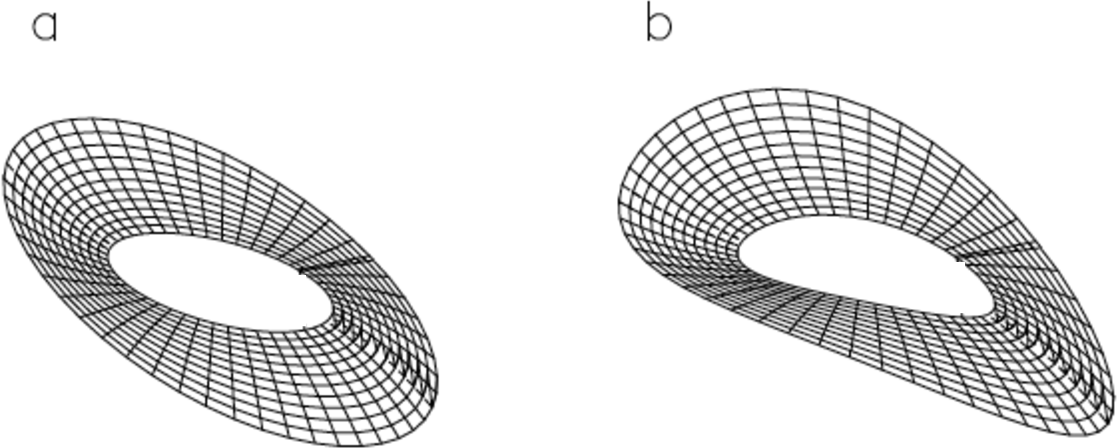}
   \includegraphics[width=0.45\textwidth]{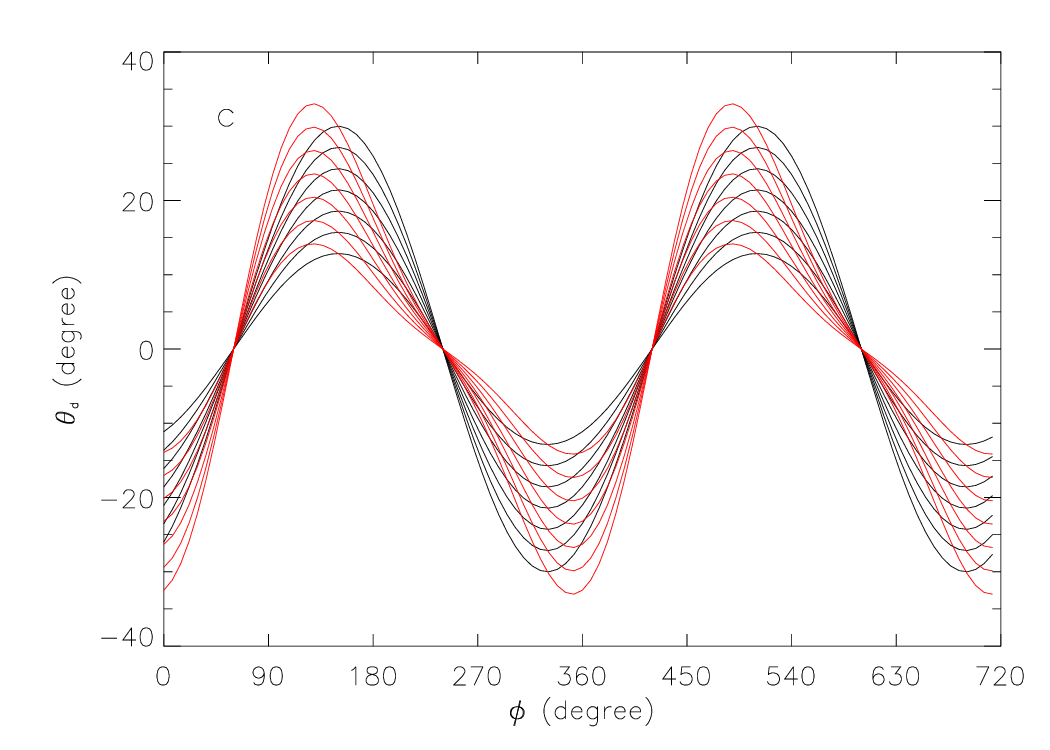}
\caption{(a): Schematic of an antisymmetric warped disk with a pure azimuthal sinusoidal function, $\sin(\phi - \phi_{\mathrm{tw}})$. (b): Similar to (a), but with an additional higher-order harmonic term, $0.25 \times \sin(2\phi - \phi_{\mathrm{tw}}^\prime)$, incorporated into the azimuthal profile, rendering the disk asymmetric. (c): Black curves: $\theta_d$ versus azimuth angle $\phi$ at different radii $r$ computed using the disk model equation from Eq~\ref{hvdisk} with the parameters shown in Figure 7 of~\citet{hic05}. Red curves: Similar to the black curves but with the higher-order harmonic term $0.25 \times \sin(2\phi - \phi_{\mathrm{tw}}^\prime)$, where $\phi_{\mathrm{tw}}^\prime = 2\pi/3$, incorporated into the azimuthal profile. \label{warpeddisk}}
\end{figure}

Conversely, the observation that the soft X-ray superorbital phases led those of the hard X-rays after MJD 57000 suggests a change in the relative azimuthal location of the reprocessing region on the warped disk. This behavior can be interpreted as a transition in the disk geometry from antisymmetric to asymmetric. From Eq.~\ref{hvdisk}, a straightforward way to break the antisymmetry is to introduce a higher-order harmonic term into the azimuthal function, namely, $\sin(\phi-\phi_{\mathrm{tw}}) \rightarrow \sin(\phi-\phi_{\mathrm{tw}})+\alpha \sin(2\phi - \phi_{\mathrm{tw}}^\prime)$,  where $\alpha$ is a real coefficient. A comparison between the antisymmetric and asymmetric disks, with an additional higher-order harmonic term of $\alpha=0.25$ in the latter case, is shown in Figures~\ref{warpeddisk}(a) and (b). The variation of $\theta_{\mathrm{d}}$ as a function of the azimuthal angle $\phi$ at different radii for both disk geometries is displayed in Figure~\ref{warpeddisk}(c). Introducing this term shifts the relative azimuthal location of the reprocessing region on the warped disk, thereby producing phase shifts between the soft and hard X-ray bands. Higher-order harmonics in the disk's azimuthal structure can naturally arise from nonlinear radiation torques, tidal interactions with the companion star, or magnetic stresses in the inner disk. These mechanisms can distort an otherwise antisymmetric warp into an asymmetric shape, thereby accounting for the observed phase shift.

Although the detected superorbital phase shift in the soft X-ray band was small (0.044 cycles = 15.8$^{\circ}$), the corresponding change in the azimuthal location of the reprocessing region was substantially larger. \citet{amb22} analyzed the broadband X-ray spectrum of LMC X-4 using Swift XRT and NuSTAR, showing that photons in the 2-50 keV band are dominated by pulsar emission and scattering processes. They further demonstrated that the soft excess at 0.6-2.0 keV originates from reprocessing in the inner accretion disk, whereas photons below 0.6 keV are primarily from thermal emission in the outer disk. In this study, we used RXTE ASM and MAXI data covering the 2-10 keV range (e.g., MAXI GSC) as the soft X-ray band. In this band, most photons still come from the pulsar, with only a minor contribution from disk reprocessing. Because this fraction is small, a modest shift in the location of the reprocessing region would have a negligible effect on the overall phase of the observed band. Therefore, to account for the observed  0.044-cycle phase shift in the 2-10 keV light curve, the azimuthal displacement of the reprocessing region must be considerably larger. If superorbital profiles could be measured at even softer energies, such as with MAXI SSC (sensitive to 0.5-12 keV), a larger phase shift would be expected because the reprocessed component would contribute more significantly. However, due to the limited effective area of MAXI SSC (200 cm$^2$), no clear superorbital profile of LMC X-4 could be obtained from these data.

\citet{bru20} analyzed LMC X-4 data obtained from simultaneous XMM-Newton and NuSTAR observations on October 30 (MJD 57325), November 4 (MJD 57330), November 11 (MJD 57337), and November 27 (MJD 57353) 2015, labeled L1-L4. These epochs corresponded to distinct superorbital phases within a superorbital cycle. For the timing analysis, the authors used NuSTAR events in the 8-60 keV range to trace hard X-ray pulsations originating from the pulsar beam, and events in the 0.5-1 keV range to monitor soft X-ray pulsations associated with reprocessing emission. Distinct pulse-phase shifts were observed among the four epochs, confirming the presence of a precessing warped disk in the LMC X-4 system.~\citet{bru20} defined the superorbital phase zero as the onset of the high state, and determined the superorbital phases by comparing the observed soft-pulse profiles with simulated profiles of the reprocessing region predicted by the warped disk model of ~\citet{hic05}. The resulting superorbital phases for L1-L4 were 0.75, 0.875, 0.125, and 0.625. When restricted to the range -0.5 to 0.5, the corresponding values became -0.25, -0.125,  0.125, and -0.375, as listed in Table~\ref{phdif}. We note that~\citet{bru18} also reported flares and pulse dropouts in LMC X-4, but these occurred during the superorbital high state and are therefore not attributable to disk absorption, in contrast to the phase shifts investigated in~\citet{bru20}.

\begin{table}
\begin{center}
  \caption{Comparison of Superorbital Phases from~\citet{bru20} and from the Sinusoidal Ephemeris in This Work.  \label{phdif}}

  \begin{tabular}{cccccc}
    \\
    \tableline\tableline
     \tableline\tableline
Observation & Date  & $\phi_B$\tablenotemark{a} & $\phi^{\prime}$\tablenotemark{b}   & $\phi_{sinu}$\tablenotemark{c} & $\Delta \phi$\tablenotemark{d}\\
 & (MJD) &          &  &             &      \\
\tableline
L1  & 57325 & -0.250 & -0.600 & -0.152 & -0.448\\
L2  & 57330 & -0.125 & -0.475 &  0.013 &  -0.488\\
L3  & 57337 & 0.125  & -0.225 & 0.243 &  -0.468\\
L4  & 57353 & -0.375 & -0.725 & -0.230 & -0.495\\
\tableline\tableline
\tableline\tableline
  \end{tabular}
      \tablenotetext{}{\parbox[t]{0.95\columnwidth}{\raggedright Notes:}}
  \tablenotetext{a}{\parbox[t]{0.9\columnwidth} {\raggedright Superorbital phase reported by~\citet{bru20}. }}
  \tablenotetext{b}{\parbox[t]{0.9\columnwidth} {\raggedright  $\phi^{\prime}=\phi_B - 0.35$ where $\phi^{\prime}$ is adjusted to align the phase zero eopch with that used in this study.} }
  \tablenotetext{c}{\parbox[t]{0.9\columnwidth} {\raggedright Superorbital phase evaluated by the sinisodial ephemeris derived in this work.}}
 \tablenotetext{d}{\parbox[t]{0.9\columnwidth} {\raggedright $\Delta \phi = \phi^{\prime} - \phi_{sinu}$}}.
\end{center}

\end{table}

In this study, we established superorbital ephemerides to characterize the phase-evolution trend, thereby providing an independent validation of the superorbital phases proposed by~\citet{bru20} for L1-L4. Because different definitions of the phase-zero epoch were used, an adjustment was necessary to align the phases. Although~\citet{bru20} did not explicitly specify the date of the phase-zero epoch, the onset of the high state appears at approximately phase -0.4 to -0.3 (as shown in the bottom panel of Figure~\ref{expprof}). By adopting the mean value of -0.35 as the onset of the high state, all phase values for L1-L4 reported by~\citet{bru20} were shifted by -0.35 and subsequently compared with those derived from the sinusoidal ephemeris (Eq.~\ref{sinueph}) at the corresponding observation dates. This comparison revealed substantial discrepancies of approximately 0.5 cycles between the two sets of phase values (see Table~\ref{phdif}). These deviations persist even after accounting for potential systematic phase fluctuations ($\pm$0.1 cycles), uncertainties in defining the start of the high state ($\pm$0.05 cycles), and the finite resolution (0.125 cycles) used by~\citet{bru20} to determine the superorbital phases. Comparable results were obtained when the cubic (Eq.~\ref{ceph}) and quartic (Eq.~\ref{4eph}) ephemerides were employed. The observed discrepancies are likely attributable to variations in the geometry of the warped disk, as all relevant observations were conducted after MJD 57000. Consequently, the antisymmetric disk model of~\citet{hic05}, adopted by~\citet{bru20}, was probably no longer applicable, given that the disk had become asymmetric. Significant deviations were therefore expected, because~\citet{bru20} evaluated the superorbital phases for L1-L4 using soft X-ray pulses reflected by the warped disk.

We identified a correlation between the superorbital phase shift and the flux variation in the hard X-ray band of LMC X-4, as demonstrated in Section~\ref{ps}. Notably, no significant flux change was detected in the soft X-ray band, suggesting that the observed effect is likely attributable to partial absorption or scattering of the neutron star's hard X-ray emission by the accretion disk. This absorption may result from changes in the disk's inclination angle or overall geometry. Such changes in the accretion disk likely have a smaller effect on the reprocessing region as seen by the observer; therefore, the flux variation in the soft X-ray band is expected to be negligible.

A similar superorbital phase shift in the soft X-ray band has been observed in Her X-1, an LMXB with a stable 35-day superorbital period. ~\citet{cla03} analyzed the RXTE ASM light curves of Her X-1 and identified a distinct superorbital phase shift of $\sim$0.1 cycles in the dynamic folded light curve~\citep[see Figure 9 in][]{cla03}, detected approximately 15 cycles before the source entered its third anomalous low state. This anomalous low state has been interpreted as obscuration of the central source by the disk~\citep{vrt01}. The hard X-ray flux probably decreases in conjunction with the superorbital phase shift, as we observed in LMC X-4. However, simultaneous hard X-ray observations were not available to directly compare with the RXTE ASM results reported by~\citet{cla03} for Her X-1. If a change in the phase difference between the soft and hard X-ray bands occurred during this phase-shift epoch of Her X-1, the disk-shape-change scenario discussed here could naturally account for the analogous phenomena observed in both LMC X-4 and Her X-1.

Our phase analysis provides strong evidence for the exceptional stability of the LMC X-4 superorbital period and for the emergence of a soft-hard X-ray phase shift after MJD 57000. However, the underlying physical mechanisms remain uncertain. Although attributing the long-term sinusoidal modulation to internal disk dynamics and extending the warped-disk model to include higher harmonics are both plausible interpretations, these scenarios are not uniquely supported by the current data. They require further testing through more extensive multiwavelength observations at softer energies, as well as detailed theoretical modeling and numerical simulations of warped accretion disks. Future campaigns utilizing next-generation X-ray facilities, combined with advances in disk-warping theory, will be crucial for elucidating the physical origins of the observed phenomena.

\section{Summary} \label{sum}

The superorbital period of LMC X-4 is among the most stable observed in X-ray binaries exhibiting superorbital modulation~\citep{ogi01}, a stability further corroborated by dynamic power spectral analyses~\citep{cla03,kot12}. However, phase analysis offers a more precise method for tracking period evolution. In this study, we analyzed the superorbital phase evolution of LMC X-4 using 33 years of all-sky X-ray monitoring data from CGRO BATSE, RXTE ASM, Swift BAT, MAXI GSC, and Fermi GBM. The phase evolution reveals a smooth long-term trend with superposed systematic fluctuations. We modeled this trend using cubic, quartic, and sinusoidal functions, finding that the quartic and sinusoidal models provided superior fits (Section~\ref{spe}) and derived the corresponding ephemerides. The sinusoidal model yielded a period of 8,898 days, representing the timescale of the smooth trend; however, this periodicity is unlikely to result from orbital motion around a hierarchical third body (Section~\ref{spv}).

The systematic phase fluctuations resemble stochastic glitches in the superorbital period, with characteristic timescales of several hundred days. Using the intervals defined by~\citet{mol15}, we estimated the rms period variation attributable to these fluctuations, which exceeds that inferred from the smooth trend (Table~\ref{rmsp}). Nevertheless, the overall rms variation of the superorbital period across the 33-year baseline is only about 0.55\%, providing the first quantitative measure of the period stability of LMC X-4. Comparable fluctuations have been reported for Her X-1 and may arise from torques acting on the warped accretion disk or from variations in the fiducial point.

A small but statistically significant phase shift of $0.044 \pm 0.010$ cycles was detected between the soft and hard X-ray bands during MJD 57000-60461, whereas no such shift was evident in the earlier epoch (MJD 53461-57000). We attribute this shift to a change in the geometry of the warped disk around MJD 57000. Because soft X-ray photons originate from reprocessing of the pulsar beam in the warped accretion disk, we modified the azimuthal function of the warped-disk model proposed by~\citet{hic05} to include a higher harmonic term, allowing the disk geometry to evolve from antisymmetric to asymmetric. This modification can account for the observed phase shift. Notably, the phase shift coincided with a decrease in hard X-ray flux, likely analogous to the anomalous low state observed in Her X-1, which is attributed to obscuration of the X-ray emission region due to variations in disk structure. We propose that geometric changes in a warped disk could similarly explain the superorbital phase shift observed in Her X-1, as evident in the dynamic folded RXTE ASM light curve reported by~\citet{cla03}.

Although we have proposed plausible scenarios to interpret the long-term sinusoidal trend, systematic phase fluctuations, and the soft-hard phase shift, these explanations remain provisional. Their physical origins are still uncertain and require confirmation through future work. In particular, coordinated multi-wavelength monitoring at softer X-ray energies , such as with the Wide-field X-ray Telescope aboard the Einstein Probe~\citep{yua15}, which is sensitive to 0.5-4.0 keV X-ray photons and surveys the entire sky over three satellite orbits, combined with advances in theoretical modeling and numerical simulations of warped accretion disks, will be essential for elucidating the mechanisms driving these behaviors in LMC X-4.


\begin{acknowledgments}
The author thanks the anonymous referee for valuable suggestions that improved the manuscript, and is grateful to Dr. Mike T. -C. Yang for his technical assistance in resolving software issues, which was essential for the data analysis. This research has made use of data and software provided by the High Energy Astrophysics Science Archive Research Center (HEASARC), which is a service of the Astrophysics Science Division at NASA/GSFC. We express our gratitude to the Gamma-Ray Astrophysics Team of the  NSSTC for archiving the CGRO BATSE and Fermi GBM data. This study also made use of data provided by RXTE ASM team at MIT and GSFC. We also acknowledge the use of public data from the Swift data archive, as well as the  MAXI data provided by RIKEN, JAXA and the MAXI team.
\end{acknowledgments}

%

\vspace{5mm}
\facilities{ADS, HEASRAC, CGRO (BATSE), RXTE (ASM), Swift (BAT), MAXI (GSC), Fermi (GBM)}
\software{heasoft (v6.35.1)}





\end{CJK*}


\begin{thebibliography}{}

\bibitem[Alcock et al.(2001)]{alc01} Alcock, C., Allsman, R.~A., Alves, D.~R., et al.\ 2001, \mnras, 321, 4, 678. doi:10.1046/j.1365-8711.2001.04041.x
\bibitem[Ambrosi et al.(2022)]{amb22} Ambrosi, E., D'A{\`\i}, A., Del Santo, M., et al.\ 2022, \mnras, 512, 3, 3422. doi:10.1093/mnras/stac450

\bibitem[Anderson et al.(1983)]{and83} Anderson, S.~F., Margon, B., \& Grandi, S.~A.\ 1983, \apj, 273, 697. doi:10.1086/161404

\bibitem[Barthelmy et al.(2005)]{bar05} Barthelmy, S.~D., Barbier, L.~M., Cummings, J.~R., et al.\ 2005, \ssr, The Burst Alert Telescope (BAT) on the SWIFT Midex Mission, 120, 3-4, 143. doi:10.1007/s11214-005-5096-3

\bibitem[Boynton et al.(1980)]{boy80} Boynton, P.~E., Crosa, L.~M., \& Deeter, J.~E.\ 1980, \apj, 237, 169. doi:10.1086/157856

 \bibitem[Bozzo et al.(2017)]{boz17} Bozzo, E., Oskinova, L., Lobel, A., et al.\ 2017, \aap, 606, L10. doi:10.1051/0004-6361/201731930
\
\bibitem[Brumback et al.(2018)]{bru18} Brumback, M.~C., Hickox, R.~C., Bachetti, M., et al.\ 2018, \apjl, 861, 1, L7. doi:10.3847/2041-8213/aacd13

\bibitem[Brumback et al.(2020)]{bru20} Brumback, M.~C., Hickox, R.~C., F{\"u}rst, F.~S., et al.\ 2020, \apj, 888, 2, 125. doi:10.3847/1538-4357/ab5b04

\bibitem[Brumback et al.(2023)]{bru23} Brumback, M.~C., Vasilopoulos, G., Coley, J.~B., et al.\ 2023, \apj, 953, 1, 89. doi:10.3847/1538-4357/ace04f

\bibitem[Chou et al.(2001)]{chou01} Chou, Y., Grindlay, J.~E., \& Bloser, P.~F.\ 2001, \apj, 549, 2, 1135. doi:10.1086/319443
  
\bibitem[Chou et al.(2008)]{chou08} Chou, Y., Chung, Y., Hu, C.-P., et al.\ 2008, \apj, 678, 2, 1316. doi:10.1086/529126
  
\bibitem[Chou(2014)]{chou14} Chou, Y.\ 2014, Research in Astronomy and Astrophysics, 14, 11, 1367-1382. doi:10.1088/1674-4527/14/11/001

\bibitem[Chou et al.(2025)]{chou25} Chou, Y., Wu, J.-L., Chen, B.-C., et al.\ 2025, \apj, 981, 1, 43. doi:10.3847/1538-4357/adaded

\bibitem[Clarkson et al.(2003)]{cla03} Clarkson, W.~I., Charles, P.~A., Coe, M.~J., et al.\ 2003, \mnras, 343, 4, 1213. doi:10.1046/j.1365-8711.2003.06761.x

\bibitem[Corbet et al.(1999)]{cor99} Corbet, R.~H.~D., Finley, J.~P., \& Peele, A.~G.\ 1999, \apj, 511, 2, 876. doi:10.1086/306727

\bibitem[Corbet et al.(2021)]{cor21} Corbet, R.~H.~D., Coley, J.~B., Krimm, H.~A., et al.\ 2021, \apj, 906, 1, 13. doi:10.3847/1538-4357/abc477

\bibitem[Dage et al.(2022)]{dag22} Dage, K.~C., Brumback, M., Neilsen, J., et al.\ 2022, \mnras, 514, 4, 5457. doi:10.1093/mnras/stac1674


\bibitem[Davison \& Fabian(1974)]{dav74} Davison, P.~J.~N. \& Fabian, A.~C.\ 1974, \mnras, 169, 27P. doi:10.1093/mnras/169.1.27P

\bibitem[Falanga et al.(2015)]{fal15} Falanga, M., Bozzo, E., Lutovinov, A., et al.\ 2015, \aap, 577, A130. doi:10.1051/0004-6361/201425191

\bibitem[Fishman et al.(1994)]{fis94} Fishman, G.~J., Meegan, C.~A., Wilson, R.~B., et al.\ 1994, \apjs, The First BATSE Gamma-Ray Burst Catalog, 92, 229. doi:10.1086/191968
  
\bibitem[Gehrels et al.(1993)]{geh93} Gehrels, N., Chipman, E., \& Kniffen, D.~A.\ 1993, \aaps, The Compton Gamma Ray Observatory., 97, 5.

\bibitem[Giacconi et al.(1972)]{gia72} Giacconi, R., Murray, S., Gursky, H., et al.\ 1972, \apj, 178, 281. doi:10.1086/151790

\bibitem[Gruber \& Rothschild(1984)]{gru84} Gruber, D.~E. \& Rothschild, R.~E.\ 1984, \apj, 283, 546. doi:10.1086/162338

\bibitem[Heemskerk \& van Paradijs(1989)]{hee89} Heemskerk, M.~H.~M. \& van Paradijs, J.\ 1989, \aap, 223, 154

\bibitem[Hickox et al.(2004)]{hic04} Hickox, R.~C., Narayan, R., \& Kallman, T.~R.\ 2004, \apj, 614, 2, 881. doi:10.1086/423928

\bibitem[Hickox \& Vrtilek(2005)]{hic05} Hickox, R.~C. \& Vrtilek, S.~D.\ 2005, \apj, 633, 2, 1064. doi:10.1086/491596

\bibitem[Hu et al.(2008)]{hu08} Hu, C.-P., Chou, Y., \& Chung, Y.-Y.\ 2008, \apj, A Parameterization Study of the Properties of the X-Ray Dips in the Low-Mass X-Ray Binary X1916-053, 680, 2, 1405. doi:10.1086/527549

\bibitem[Hu et al.(2015)]{hu15} Hu, C.-P., Lin, C.-P., Chou, Y., et al.\ 2015, Publication of Korean Astronomical Society, On the Complex Variability of the Superorbital Modulation Period of LMC X-4, 30, 2, 595. doi:10.5303/PKAS.2015.30.2.595
\bibitem[Hung et al.(2010)]{hun10} Hung, L.-W., Hickox, R.~C., Boroson, B.~S., et al.\ 2010, \apj, 720, 2, 1202. doi:10.1088/0004-637X/720/2/1202  

\bibitem[Iaria et al.(2015)]{iar15} Iaria, R., Di Salvo, T., Gambino, A.~F., et al.\ 2015, \aap, 582, A32. doi:10.1051/0004-6361/201526500

\bibitem[Ilovaisky et al.(1984)]{ilo84} Ilovaisky, S.~A., Chevalier, C., Motch, C., et al.\ 1984, \aap, 140, 251. 

\bibitem[Islam et al.(2023)]{isl23} Islam, N., Corbet, R.~H.~D., Coley, J.~B., et al.\ 2023, \apj, 948, 1, 45. doi:10.3847/1538-4357/acbc19
  
\bibitem[Jain et al.(2024)]{jai24} Jain, C., Sharma, R., \& Paul, B.\ 2024, \mnras, 529, 4, 4056. doi:10.1093/mnras/stae784

\bibitem[Krimm et al.(2013)]{kri13} Krimm, H.~A., Holland, S.~T., Corbet, R.~H.~D., et al.\ 2013, \apjs, 209, 1, 14. doi:10.1088/0067-0049/209/1/14

\bibitem[Kelley et al.(1983a)]{kel83a} Kelley, R.~L., Jernigan, J.~G., Levine, A., et al.\ 1983, \apj, 264, 568. doi:10.1086/160626
  \
\bibitem[Kelley et al.(1983b)]{kel83b} Kelley, R.~L., Rappaport, S., Clark, G.~W., et al.\ 1983, \apj, 268, 790. doi:10.1086/161001

\bibitem[Koenigsberger et al.(2006)]{koe06} Koenigsberger, G., Georgiev, L., Moreno, E., et al.\ 2006, \aap, 458, 2, 513. doi:10.1051/0004-6361:20065305

\bibitem[Kotze \& Charles(2012)]{kot12} Kotze, M.~M. \& Charles, P.~A.\ 2012, \mnras, 420, 2, 1575. doi:10.1111/j.1365-2966.2011.20146.x

\bibitem[Kulkarni \& Romanova(2013)]{kul13} Kulkarni, A.~K. \& Romanova, M.~M.\ 2013, \mnras, 433, 4, 3048. doi:10.1093/mnras/stt945

\bibitem[Lang et al.(1981)]{lan81} Lang, F.~L., Levine, A.~M., Bautz, M., et al.\ 1981, \apjl, 246, L21. doi:10.1086/183545 

\bibitem[Levine et al.(1991)]{lev91} Levine, A., Rappaport, S., Putney, A., et al.\ 1991, \apj, 381, 101. doi:10.1086/170632

\bibitem[Levine et al.(1996)]{lev96} Levine, A.~M., Bradt, H., Cui, W., et al.\ 1996, \apjl, First Results from the All-Sky Monitor on the Rossi X-Ray Timing Explorer, 469, L33. doi:10.1086/310260

\bibitem[Levine et al.(2000)]{lev00} Levine, A.~M., Rappaport, S.~A., \& Zojcheski, G.\ 2000, \apj, 541, 1, 194. doi:10.1086/309398

\bibitem[Levine et al.(2011)]{lev11} Levine, A.~M., Bradt, H.~V., Chakrabarty, D., et al.\ 2011, \apjs, An Extended and More Sensitive Search for Periodicities in Rossi X-Ray Timing Explorer/All-Sky Monitor X-Ray Light Curves, 196, 1, 6. doi:10.1088/0067-0049/196/1/6

\bibitem[Li et al.(1978)]{li78} Li, F., Rappaport, S., \& Epstein, A.\ 1978, \nat, 271, 37. doi:10.1038/271037a0

\bibitem[Margon \& Anderson(1989)]{mar89} Margon, B. \& Anderson, S.~F.\ 1989, \apj, 347, 448. doi:10.1086/168132

\bibitem[Martin \& Charles(2024)]{mar24} Martin, R.~G. \& Charles, P.~A.\ 2024, \mnras, 528, 1, L59. doi:10.1093/mnrasl/slad170

\bibitem[Matsuoka et al.(2009)]{mat09} Matsuoka, M., Kawasaki, K., Ueno, S., et al.\ 2009, \pasj, The MAXI Mission on the ISS: Science and Instruments for Monitoring All-Sky X-Ray Images, 61, 999. doi:10.1093/pasj/61.5.999 

\bibitem[Meegan et al.(1992)]{mee92} Meegan, C.~A., Fishman, G.~J., Wilson, R.~B., et al.\ 1992, \nat, Spatial distribution of {\ensuremath{\gamma}}-ray bursts observed by BATSE, 355, 6356, 143. doi:10.1038/355143a0

\bibitem[Meegan et al.(2009)]{mee09} Meegan, C., Lichti, G., Bhat, P.~N., et al.\ 2009, \apj, The Fermi Gamma-ray Burst Monitor, 702, 1, 791. doi:10.1088/0004-637X/702/1/791

\bibitem[Molkov et al.(2015)]{mol15} Molkov, S.~V., Lutovinov, A.~A., \& Falanga, M.\ 2015, Astronomy Letters, Determination of parameters of Long-Term variability of the X-ray pulsar LMC X-4, 41, 10, 562. doi:10.1134/S1063773715100047

\bibitem[Molkov et al.(2017)]{mol17} Molkov, S., Lutovinov, A., Falanga, M., et al.\ 2017, \mnras, 464, 2, 2039. doi:10.1093/mnras/stw2429

\bibitem[Naik \& Paul(2004)]{nai04} Naik, S. \& Paul, B.\ 2004, \apj, 600, 1, 351. doi:10.1086/379803

\bibitem[Naranan et al.(1985)]{nar85} Naranan, S., Elsner, R.~F., Darbro, W., et al.\ 1985, \apj, 290, 487. doi:10.1086/163006

\bibitem[Ogilvie \& Dubus(2001)]{ogi01} Ogilvie, G.~I. \& Dubus, G.\ 2001, \mnras, 320, 4, 485. doi:10.1046/j.1365-8711.2001.04011.x

\bibitem[Patruno \& Watts(2021)]{pat21} Patruno, A. \& Watts, A.~L.\ 2021, Timing Neutron Stars: Pulsations, Oscillations and Explosions, 461, 143. doi:10.1007/978-3-662-62110-3\_4

\bibitem[Paul \& Kitamoto(2002)]{pau02} Paul, B. \& Kitamoto, S.\ 2002, Journal of Astrophysics and Astronomy, 23, 1-2, 33. doi:10.1007/BF02702462

\bibitem[Petterson(1977)]{pet77} Petterson, J.~A.\ 1977, \apj, 218, 783. doi:10.1086/155735

\bibitem[Pietsch et al.(1985)]{pie85} Pietsch, W., Pakull, M., Voges, W., et al.\ 1985, \ssr, 40, 3-4, 371. doi:10.1007/BF00179843

\bibitem[Pringle(1996)]{pri96} Pringle, J.~E.\ 1996, \mnras, 281, 1, 357. doi:10.1093/mnras/281.1.357

 \bibitem[Priedhorsky \& Terrell(1984)]{pri84} Priedhorsky, W. \& Terrell, J.\ 1984, \apjl, 284, L17. doi:10.1086/184343
 
\bibitem[Ransom et al.(2014)]{ran14} Ransom, S.~M., Stairs, I.~H., Archibald, A.~M., et al.\ 2014, \nat, 505, 7484, 520. doi:10.1038/nature12917

\bibitem[Safi Harb et al.(1996)]{saf96} Safi Harb, S., Ogelman, H., \& Dennerl, K.\ 1996, \apjl, 456, L37. doi:10.1086/309859

\bibitem[Staubert et al.(2009)]{sta09} Staubert, R., Klochkov, D., Postnov, K., et al.\ 2009, \aap, 494, 3, 1025. doi:10.1051/0004-6361:200810743

\bibitem[Staubert et al.(2013)]{sta13} Staubert, R., Klochkov, D., Vasco, D., et al.\ 2013, \aap, 550, A110. doi:10.1051/0004-6361/201220316

\bibitem[Trowbridge et al.(2007)]{tro07} Trowbridge, S., Nowak, M.~A., \& Wilms, J.\ 2007, \apj, 670, 1, 624. doi:10.1086/522075

\bibitem[Vrtilek et al.(2001)]{vrt01} Vrtilek, S.~D., Quaintrell, H., Boroson, B., et al.\ 2001, \apj, 549, 1, 522. doi:10.1086/319071

\bibitem[Wen et al.(2006)]{wen06} Wen, L., Levine, A.~M., Corbet, R.~H.~D., et al.\ 2006, \apjs, 163, 2, 372. doi:10.1086/500648

\bibitem[Wijers \& Pringle(1999)]{wij99} Wijers, R.~A.~M.~J. \& Pringle, J.~E.\ 1999, \mnras, 308, 1, 207. doi:10.1046/j.1365-8711.1999.02720.x

\bibitem[Wilson-Hodge et al.(2012)]{wil12} Wilson-Hodge, C.~A., Case, G.~L., Cherry, M.~L., et al.\ 2012, \apjs, Three Years of Fermi GBM Earth Occultation Monitoring: Observations of Hard X-Ray/Soft Gamma-Ray Sources, 201, 2, 33. doi:10.1088/0067-0049/201/2/33

\bibitem[Wojdowski et al.(1998)]{woj98} Wojdowski, P., Clark, G.~W., Levine, A.~M., et al.\ 1998, \apj, 502, 1, 253. doi:10.1086/305893

\bibitem[Woo et al.(1996)]{woo96} Woo, J.~W., Clark, G.~W., Levine, A.~M., et al.\ 1996, \apj, 467, 811. doi:10.1086/177655

\bibitem[Yuan \& Osborne(2015)]{yua15} Yuan, W. \& Osborne, J.~P.\ 2015, , arXiv:1506.07736. doi:10.48550/arXiv.1506.07736

  
\end{thebibliography}
\end{document}